\documentclass[amsfonts,amsmath,prd,preprint,nofootinbib]{revtex4}
\newcommand{\beq}{\begin{equation}}
\newcommand{\eeq}{\end{equation}}

\newcommand{\bsp}{\begin{split}}

\usepackage{epsfig,bbm,cancel,ulem}
\usepackage{color}
\usepackage[breaklinks=true]{hyperref}

\begin{document}

\title{Charged black holes in higher-dimensional Eddington-inspired Born-Infeld gravity}

\author{Byon N. Jayawiguna}
\email{byon.nugraha@ui.ac.id}
\author{Handhika S. Ramadhan}
\email{hramad@ui.ac.id}
\affiliation{Departemen Fisika, FMIPA, Universitas Indonesia, Depok, 16424, Indonesia. }
\def\changenote#1{\footnote{\bf #1}}

\begin{abstract}
We study static (electrically)-charged solutions of Eddington-inspired Born-Infeld (EiBI) theory of gravity in general $D$-dimensional spacetime. We consider both linear (Maxwell) as well as nonlinear electrodynamics for the matter fields. In this particular work, the nonlinear theory we specifically consider is the Born-Infeld (BI) electrodynamics. The solutions describe higher-dimensional black holes in EiBI gravity. For the linear Maxwell field, we show that the electric field is still singular for $D>4$. This singularity is cured when EiBI is coupled to the BI electrodynamics. We obtain EiBI-BI black hole solutions in the limit of ${\tilde\alpha}\equiv4\kappa b^2/\lambda=1$ and $2$. We also investigate their thermodynamical property. We show that all solutions satisfy the first-law of black hole thermodynamics, from which their corresponding ADM mass can be extracted. It is found that $\kappa$ imposes a charge screening that makes the corresponding Hawking temperature experiences some sudden jump from charged-type to the Schwarzschild-type at some critical value of $\kappa$. Thermodynamical stability reveals that the EiBI-BI black holes can exist with smaller horizon than their Reissner-Nordstrom (RN) counterparts. 

\end{abstract}

\maketitle
\thispagestyle{empty}
\setcounter{page}{1}

\section{Introduction}
\label{sec:introduction}

Black hole has been an intriguing phenomenon in gravitational physics that fascinates both theoretical physicists as well as astronomers. In General Relativity (GR), the theoretical existence of black holes is inevitable. Observationally, the recent detection of gravitational waves is a solid proof that such object does exist physically~\cite{Abbott:2016blz}.

Although as a modern cosmological framework general relativity is very successful, several large-scale  or strong-field regime phenomena (such as the accelerated expansion of the universe, the big bang singularity, {\it etc}) demands a more satisfactory explanation. This is the main reason behind the vast modern literature on modified-gravitational theories beyond GR.  Among many excellent alternative theory of gravity, the Eddington-inspired Born-Infeld (EiBI) theory proposed by Banados and Ferreira recently enjoys widespread attention~\cite{Banados:2010ix}. Constructed based on the old proposal of Eddington gravitational action~\cite{eddington} combined with the nonlinearity of the Born-Infeld (BI) theory~\cite{Born:1934gh}, this particular theory resurrects the popular old-Born-Infeld gravity models~\cite{Deser:1998rj, Nieto:2004qj}. This theory offers interesting solutions to some theoretical and cosmological problems, like the freedom from ghosts and instabilities~\cite{Delsate:2012ky}, or the non-singularity of big bang and big crunch~\cite{Avelino:2012ue, Pani:2011mg} (however, see the discussions in~\cite{Pani:2012qd, Shaikh:2018cul, Li:2017ttl}). In nuclear astrophysics the EiBI model finds its (perhaps) most active elaboration since coupling it to the NS equation of state (EOS) enables one to obtain the observed mass of the Neutron Star (NS) while simultaneously solves the ``hyperon puzzle" (see, for example,~\cite{Qauli:2016vza, Qauli:2017ntr} and references therein). For a comprehensive review on EiBI gravity, see~\cite{BeltranJimenez:2017doy}.

Despite the extensive investigation on the astrophysical and cosmological aspects of EiBI, its black hole aspects are less studied. The simplest black hole solution in this theory is trivial: the Schwarzschild-(A)dS solution. This is because the corresponding EiBI action reduces to the ordinary Einstein-Hilbert in the vacuum. The known non-trivial black holes in EiBI gravity, the electrically-charged ones, are discussed in~\cite{Banados:2010ix, Sotani:2014lua, Wei:2014dka}. Their solutions reduced to the well-known Reissner-Nordstrom-(A)dS (RN-(A)dS) in the limit of $\kappa\rightarrow0$. Jana and Kar studied the black hole with nonlinear charged by coupling EiBI with BI electrodynamics~\cite{Jana:2015cha}. Their black holes are parametrized by two parameters $\kappa$ and $b^2$ that control the nonlinearity of the gravity or the electrodynamics field, respectively. These black holes are none other than the EiBI version of the Einstein-BI black holes~\cite{Dey:2004yt, Cai:2004eh}. In the limit $\kappa\rightarrow0$ (while keeping $b^2$ finite) the solutions reduce to that of geonic black hole~\cite{Demianski:1986wx}. Recently there has also been a discussion on coupling the gravity to topological defects to produce black holes with global monopole~\cite{Lambaga:2018yzv} or to model the Lorentzian wormhole~\cite{Shaikh:2015oha, Shaikh:2018yku}, while the problem of overchanging extremal black hole and its cosmic cencorship was addressed recently in~\cite{Jana:2018knq}. 

On the other hand, from different point of view, there has also been a vast literature in the study of black hole, both in GR as well as in the alternative gravitational theories, with nonlinear electrodynamics (NLED) charge~\cite{Hendi:2010zz, Hendi:2010kv, Hendi:2010dz, Rodrigues:2015ayd, Kruglov:2015yua, Kruglov:2018rrm, Hendi:2013mka, Nojiri:2017kex, Hendi:2017oka, Hendi:2017ptl, EslamPanah:2017yoc, Hendi:2017mgb, Hendi:2017uly, Hendi:2015oda}, most of which are done in four dimensions. In this work we devote our effort to studying the generalization of charged black holes in arbitrary dimensions, both with linear (Maxwell) as well as nonlinear electrodynamics (NLED), in the same way Reissner-Nordstrom-(A)dS is generalized to higher dimensions in~\cite{Cardoso:2004uz}. The particular NLED we choose is the BI electrodynamics. Our work is organized as follows. In Section~\ref{sec:action} we briefly lay down the theoretical framework of the EiBI gravity. In Sections~\ref{sec:vacuum} we show that there is a constraint for the vacuum EiBI-(A)dS solutions to exist. Sections~\ref{sec:maxwell} and~\ref{sec:bieldin} are devoted to solving the charged black holes in both linear (Maxwell) and nonlinear(BI) electrodynamics, respectively. In Section~\ref{sec:thermo} we discuss their thermodynamical properties and the corresponding stability. Finally, we summarize our work in Section~\ref{sec:conclu}. 
 
\section{The Action}
\label{sec:action}

The action of EiBI theory in $D$ dimensions is given by~\cite{Banados:2010ix}:
\begin{equation}
\label{1}
S(g,\Gamma,\Phi)=\frac{1}{8\pi\kappa}\int d^Dx \left(\sqrt{-|g_{\mu\nu}+\kappa R_{\mu\nu}(\Gamma)|} - \lambda \sqrt{-|g_{\mu\nu}|}\right)+ S_M(g,\Phi),
\end{equation}
where we set $c=G=1$. Following Vollick~\cite{Vollick:2003qp} in this theory we employ Palatini formalism; that is, we treat the connection $\Gamma$ and the metric $g$ as two independent fields, while the Ricci tensor $R_{\mu\nu}(\Gamma)$ is built from the connection. The parameter $\lambda\equiv1+\kappa\Lambda$ corresponds to the cosmological constant, and $S_M(g_{\mu\nu},\Phi)$ is the action of matter and coupled only to the metric.Variation with respect to $\Gamma$ yields
\begin{equation}
\label{3}
\Gamma^{\sigma}_{\alpha\beta}=\frac{1}{2}q^{\rho\sigma}\left(\partial_{\alpha}q_{\beta \rho}+\partial_{\beta}q_{\rho\alpha}-\partial_{\rho} q_{\alpha\beta}\right),
\end{equation}
where 
\begin{equation}
\label{2}
q_{\mu\nu}\equiv g_{\mu\nu}+\kappa R_{\mu\nu}
\end{equation}
is the {\it auxiliary} metric. On the other hand, variation with respect to $g_{\mu\nu}$ results in
\begin{equation}
\label{4}
\sqrt{-q}q^{\mu\nu}=\lambda \sqrt{-g}g^{\mu\nu}-8\pi\kappa \sqrt{-g}T^{\mu\nu}.
\end{equation}

For the metrics, we assume spherical symmetry and and staticity~\cite{Sotani:2014lua},\cite{Wei:2014dka}, {\it i.e.,} they can be written as
\begin{eqnarray}
g_{\mu\nu}dx^{\mu}dx^{\nu}&=&-\psi^2(r)f(r)dt^2+\frac{1}{f(r)} dr^2+r^2d\Omega^2_{D-2},\label{5}\\
q_{\mu\nu}dx^{\mu}dx^{\nu}&=&-G^2(r)F(r)dt^2+\frac{1}{F(r)} dr^2+H^2(r)d\Omega^2_{D-2}.\label{6}
\end{eqnarray}
The geometrical part of the EiBI equations, Eq.~\eqref{2} \textcolor{red}{with \eqref{4}}, reads
\begin{equation}
\label{33}
\begin{split}
\frac{2G''}{G}+\frac{3G'F'}{GF}+\frac{F''}{F}+\frac{2(D-2)G'H'}{GH}+\frac{(D-2)F'H'}{FH}  \\ 
=\frac{2}{\kappa F}\left[\frac{1}{\left(\lambda-8\pi \kappa T^{0}_{0}\right)^{\frac{4-D}{D-2}} \left(\lambda+8\pi\kappa T^{0}_{0}\right)}-1\right]
\end{split}
\end{equation}
\begin{equation}
\label{34}
\begin{split}
\frac{2G''}{G}+\frac{3G'F'}{GF}+\frac{F''}{F}+\frac{2(D-2)H''}{H}+\frac{(D-2)F'H'}{FH}  \\ 
=\frac{2}{\kappa F}\left[\frac{1}{\left(\lambda-8\pi \kappa T^{1}_{1}\right)^{\frac{4-D}{D-2}} \left(\lambda+8\pi\kappa T^{1}_{1}\right)}-1\right]
\end{split}
\end{equation}
\begin{equation}
\label{35}
\begin{split}
\frac{G'H'}{GH}+\frac{F'H'}{FH}+\frac{H''}{H}+\frac{(D-3)H'^2}{H^2}-\frac{(D-3)}{FH^2} \\
=\frac{1}{\kappa F}\left[\frac{1}{\left(\lambda-8\pi \kappa T^{2}_{2}\right)^{\frac{2}{D-2}}}-1\right].
\end{split}
\end{equation}
In the following Sections we shall discuss the solutions with several $T^{\mu\nu}$.

\section{Higher-dimensional EiBI in the Vacuum}
\label{sec:vacuum}

As a warm up, let us start with EiBI black hole in the vacuum. At first it seems a pointless exercise, since we are aware that in the vacuum the EiBI is identical to GR~\cite{Banados:2010ix}. We may deduce that the solution is nothing but the Tangherlini-(A)dS metric~\cite{Tangherlini:1963bw}. However, a closer examination reveals some additional constraint on the existence of the AdS black hole in $D>4$, which does not appear in its four-dimensional counterpart.

From Eq.~\eqref{4}, with $T^{\mu\nu}=0$, we have
\begin{equation}
\label{7}
\frac{H^{D-2}}{GF}=\frac{\lambda r^{D-2}}{\psi f},
\end{equation}
\begin{equation}
\label{8}
GH^{D-2} F=\lambda \psi r^{D-2}f,
\end{equation}
\begin{equation}
\label{9}
GH^{D-4}=\lambda\psi r^{D-4}.
\end{equation}
Combining~\eqref{7}-\eqref{8}, it is trivial to show 
\begin{equation}
\label{10}
H=\lambda^{\frac{1}{D-2}} r,
\end{equation}
\begin{equation}
\label{11}
F=\frac{f}{\lambda^{\frac{2}{D-2}}},
\end{equation}
\begin{equation}
\label{12}
G= \psi \lambda^{\frac{2}{D-2}}.
\end{equation}
Also, the geometrical part of the EiBI equations, Eq.~\eqref{2}, reads
\begin{eqnarray}
\frac{2G''}{G}+\frac{3G'F'}{GF}+\frac{F''}{F}+\frac{2(D-2)G'H'}{GH}+\frac{(D-2)F'H'}{FH}=\frac{2}{\kappa F}\left(\frac{1}{\lambda^{\frac{2}{D-2}}}-1\right)\label{13},\\
\frac{2G''}{G}+\frac{3G'F'}{GF}+\frac{F''}{F}+\frac{2(D-2)H''}{H}+\frac{(D-2)F'H'}{FH}=\frac{2}{\kappa F}\left(\frac{1}{\lambda^{\frac{2}{D-2}}}-1\right)\label{14},\\
\frac{G'H'}{GH}+\frac{F'H'}{FH}+\frac{H''}{H}+\frac{(D-3)H'^2}{H^2}-\frac{(D-3)}{FH^2}=\frac{1}{\kappa F}\left(\frac{1}{\lambda^{\frac{2}{D-2}}}-1\right)\label{15}.
\end{eqnarray}
On the other hand, Eqs.~\eqref{13} and \eqref{14} can be combined to obtain $G=C_{1} H'$ which, when inserted into~\eqref{10}-\eqref{12} yields a trivial solution $\psi(r)=1$. The remaining constraint, Eq.~\eqref{15}, gives
\begin{eqnarray}
\label{17}
F(r)=\frac{1}{H'(H^{D-2})'}&& \bigg[ C_2+ \int \bigg[H' H^{D-4}(D-2)(D-3) \nonumber \\ && -\frac{(D-2)H^{D-4} H' H^2}{\kappa}\left(\frac{1}{\lambda^{\frac{2}{D-2}}}-1\right)\bigg]dr\bigg].
\end{eqnarray}
The integral is elementary, and the result enables us to get physical metric $f(r)$ from \eqref{11}, 
\begin{equation}
\label{20}
f(r)= 1-\frac{2M}{r^{D-3}} -\frac{1}{D-1}\left(\frac{\lambda^{\frac{2}{D-2}}-1}{\kappa}\right)r^2.
\end{equation}
Not surprisingly, the solution is nothing but Tangherlini-(A)dS. However, for $D>4$ we notice that the metric is real for the following conditions: 

\paragraph{dS : $\lambda= 1+ \kappa \Lambda $}   

\begin{align}
	\label{21}
	 \lambda  >1 ,  \Lambda>0 
\end{align}

\paragraph{AdS : $\lambda=1-\kappa|\Lambda|$}	

\begin{align}
	\label{22}
 \lambda & \begin{cases}
		\geq 0, & -\frac{1}{\kappa}\leq\Lambda<0 \\
		=0, & \Lambda=- \frac{1}{\kappa}
	\end{cases}.
\end{align}
Note that $ \lambda<0 $ is a forbidden case because the metric becomes complex. Therefore, $\Lambda$ and $\kappa$ (assuming $\kappa>0$) should satisfy the constraint $\Lambda<-\frac{1}{\kappa}$. 



\section{BORN-INFELD GRAVITY COUPLED TO MAXWELL ELECTRODYNAMICS IN HIGHER DIMENSIONS}
\label{sec:maxwell}

Here we consider our first non-trivial case, the Einstein-Maxwell theory in higher dimensions. The matter Lagrangian is 
\begin{equation}
\label{23}
\mathcal{L}_{Matter}=-\frac{1}{16\pi}F_{\mu\nu}F^{\mu\nu},
\end{equation}
with energy-momentum tensor
\begin{equation}
\label{24}
T_{\mu\nu}=\frac{1}{4\pi}\left(F_{\mu\sigma}F^{\sigma}_{\nu}-\frac{1}{4}g_{\mu\nu}F_{\sigma\rho}F^{\sigma\rho}\right).
\end{equation}

The matter equation is simply
\begin{equation}
\label{25}
\nabla_{\mu}F^{\mu\nu}=0.
\end{equation}
Assuming a static electrically-charged source, $A^{\mu}=\phi(r)\delta^{\mu0}$, it is easy to see that the electric field is given by
\begin{equation}
\label{27}
E =\frac{q}{r^{D-2}}\psi(r),
\end{equation}
with $q$ an integration constant which can usually be identified as the charge. The energy-momentum tensor thus yields
\begin{equation}
\label{26}
T^{\mu}_{\nu}=\frac{q^2}{8\pi r^{2(D-2)}}diag(-1,-1,1,1,.....,1).
\end{equation}

Inserting these into Eq.~\eqref{4}, we have
\begin{equation}
\label{28}
\frac{H^{D-2}}{GF}=\frac{r^{D-2}}{\psi f}\left(\lambda+\frac{\kappa q^{2}}{r^{2D-4}}\right),
\end{equation}
\begin{equation}
\label{29}
GH^{D-2}F=\psi f r^{D-2} \left(\lambda+\frac{\kappa q^{2}}{r^{2D-4}}\right),
\end{equation}
\begin{equation}
\label{30}
G=\psi \left(\lambda+\frac{\kappa q^{2}}{r^{2D-4}}\right)^{\frac{4-D}{D-2}} \left(\lambda-\frac{\kappa q^{2}}{r^{2D-4}}\right).
\end{equation}
Solving the first two equations, we get
\begin{equation}
\label{31}
H=r\left(\lambda+\frac{\kappa q^{2}}{r^{2D-4}}\right)^{\frac{1}{D-2}},
\end{equation}
\begin{equation}
\label{32}
F=f \left(\lambda+\frac{\kappa q^{2}}{r^{2D-4}}\right)^{\frac{D-4}{D-2}} \left(\lambda-\frac{\kappa q^{2}}{r^{2D-4}}\right)^{-1}.
\end{equation}

Meanwhile, Eq.~\eqref{2} reduces to
\begin{equation}
\label{33}
\begin{split}
\frac{2G''}{G}+\frac{3G'F'}{GF}+\frac{F''}{F}+\frac{2(D-2)G'H'}{GH}+\frac{(D-2)F'H'}{FH}  \\ 
=\frac{2}{\kappa F}\left[\frac{1}{\left(\lambda+\frac{\kappa q^{2}}{r^{2D-4}}\right)^{\frac{4-D}{D-2}} \left(\lambda-\frac{\kappa q^{2}}{r^{2D-4}}\right)}-1\right]
\end{split}
\end{equation}
\begin{equation}
\label{34}
\begin{split}
\frac{2G''}{G}+\frac{3G'F'}{GF}+\frac{F''}{F}+\frac{2(D-2)H''}{H}+\frac{(D-2)F'H'}{FH} \\ 
=\frac{2}{\kappa F}\left[\frac{1}{\left(\lambda+\frac{\kappa q^{2}}{r^{2D-4}}\right)^{\frac{4-D}{D-2}} \left(\lambda-\frac{\kappa q^{2}}{r^{2D-4}}\right)}-1\right]
\end{split}
\end{equation}
\begin{equation}
\label{35}
\begin{split}
\frac{G'H'}{GH}+\frac{F'H'}{FH}+\frac{H''}{H}+\frac{(D-3)H'^2}{H^2}-\frac{(D-3)}{FH^2} \\
=\frac{1}{\kappa F}\left[\frac{1}{\left(\lambda+\frac{\kappa q^2}{r^{2(D-2)}}\right)^{\frac{2}{D-2}}}-1\right].
\end{split}
\end{equation}
Combining the first two to obtain $G=C_{1} H'$, this can be substituted to \eqref{30}, along with \eqref{31}, to obtain
$\psi(r) = \frac{C_{1} r^2}{{(\lambda r^{2(D-2)}+\kappa q^2)^{\frac{1}{D-2}}}}$.
The constant $C_1$ can be determined by looking at $r \rightarrow \infty$, that $E\rightarrow\frac{q}{r^{D-2}}$. Thus, $C_{1}=\lambda^{\frac{1}{D-2}}$. Therefore,
\begin{eqnarray}
\psi(r)&=&{r^2\over\left(r^{2\left(D-2\right)}+\kappa q^2/\lambda\right)^{1\over D-2}}\label{36a},\\
E(r)&=&{q\over r^{D-4}\left(r^{2\left(D-2\right)}+\kappa q^2/\lambda\right)^{1\over D-2}}\label{37a}.
\end{eqnarray}

The metric solution can be obtained from~\eqref{35},
\begin{eqnarray}
\label{38}
F(r) = \frac{1}{H'^2 H^{D-3}} \bigg[ \frac{C_{2}}{(D-2)} +  \int\left( (D-3)H'H^{D-4}-\frac{H^{D-2}H'}{\kappa}\left(1-\frac{1}{(\lambda+\frac{\kappa q^2}{r^{2(D-2)}})^\frac{2}{D-2}} \right)  \right)\ dr \bigg],\nonumber\\
\end{eqnarray}
by inserting~\eqref{31}. The result can then be substituted into~\eqref{32} to obtain
\begin{eqnarray}
\label{39}
f(r)&=&\frac{r^{D-3}(\lambda r^{2(D-2)}+\kappa q^2)^{\frac{1}{D-2}} C_{2}}{(\lambda r^{2(D-2)}-\kappa q^2)(D-2)}\nonumber\\
&&+\frac{r^{-2D}(\lambda r^{2D} + \kappa q^{2} r^{4})\kappa q^{2}(D-3)}{(\lambda r^{2(D-2)}-\kappa q^2)\lambda (D-1)} \, _{2}F_{1}\left(1,\frac{7-3D}{4-2D};\frac{1}{2}\left(\frac{3D-5}{D-2}\right);-\frac{\kappa q^2 r^{4-2D}}{\lambda}\right)\nonumber\\
&&+\frac{r^{2-2D}(\lambda r^{2D} + \kappa q^{2} r^{4}) q^2}{(\lambda r^{2(D-2)}-\kappa q^2)\lambda(D-3)} \, _{2}F_{1}\left(1,\frac{3(D-3)}{2(D-2)};\frac{7-3D}{4-2D};-\frac{\kappa q^2 r^{4-2D}}{\lambda}\right)\nonumber\\
&&+\frac{r^{-2}(\lambda r^{2D} + \kappa q^{2} r^{4})}{(\lambda r^{2(D-2)}-\kappa q^2){(D-1)\kappa}} \, _{2}F_{1}\left(1,\frac{D-5}{2(D-2)};\frac{D-3}{2(D-2)};-\frac{\kappa q^2 r^{4-2D}}{\lambda}\right)\nonumber\\
&&+\frac{r^{-4}(\lambda r^{2D} + \kappa q^{2} r^{4})}{(\lambda r^{2(D-2)}-\kappa q^2)} \, _{2}F_{1}\left(1,\frac{D-3}{2(D-2)};\frac{D-1}{2(D-2)};-\frac{\kappa q^2 r^{4-2D}}{\lambda} \right)\nonumber \\ &&-\frac{r^{-6} (\lambda r^{2D} + \kappa q^{2} r^{4})(\lambda r^{2D-4} + \kappa q^2 )^{\frac{2}{D-2}}}{ (\lambda r^{2(D-2)}-\kappa q^2)(D-1)\kappa},
\end{eqnarray}
where $\, _2F_1\left(a,b;c;d\right)$ is the hypergeometric function~\cite{abramowitz}, and $C_2$ a constant of integration. Upon setting $q=0$ it reduces considerably to $f(r)=1 + \frac{C_{2}}{\lambda^{\frac{D-3}{D-2}} (D-2) r^{D-3} }-\frac{1}{D-1}\left(\frac{\lambda^{\frac{2}{D-2}}-1}{\kappa}\right)r^{2}$. Thus, $C_{2}=-2M\lambda^{\frac{D-3}{D-2}}(D-2)$. The metric function $f(r)$ then reads 
\begin{eqnarray}
\label{42}
f(r)&=&-\frac{r^{D-3}(\lambda r^{2(D-2)}+\kappa q^2)^{\frac{1}{D-2}} 2M\lambda^{\frac{D-3}{D-2}}}{(\lambda r^{2(D-2)}-\kappa q^2)}\nonumber\\
&&+\frac{r^{-2D}(\lambda r^{2D} + \kappa q^{2} r^{4})\kappa q^{2}(D-3)}{(\lambda r^{2(D-2)}-\kappa q^2)\lambda (D-1)} \, _{2}F_{1}\left(1,\frac{7-3D}{4-2D};\frac{1}{2}\left(\frac{3D-5}{D-2}\right);-\frac{\kappa q^2 r^{4-2D}}{\lambda}\right)\nonumber\\
&&+\frac{r^{2-2D}(\lambda r^{2D} + \kappa q^{2} r^{4}) q^2}{(\lambda r^{2(D-2)}-\kappa q^2)\lambda(D-3)} \, _{2}F_{1}\left(1,\frac{3(D-3)}{2(D-2)};\frac{7-3D}{4-2D};-\frac{\kappa q^2 r^{4-2D}}{\lambda}\right)\nonumber\\
&&+\frac{r^{-2}(\lambda r^{2D} + \kappa q^{2} r^{4})}{(\lambda r^{2(D-2)}-\kappa q^2){(D-1)\kappa}} \, _{2}F_{1}\left(1,\frac{D-5}{2(D-2)};\frac{D-3}{2(D-2)};-\frac{\kappa q^2 r^{4-2D}}{\lambda}\right)\nonumber\\
&&+\frac{r^{-4}(\lambda r^{2D} + \kappa q^{2} r^{4})}{(\lambda r^{2(D-2)}-\kappa q^2)} \, _{2}F_{1}\left(1,\frac{D-3}{2(D-2)};\frac{D-1}{2(D-2)};-\frac{\kappa q^2 r^{4-2D}}{\lambda} \right)\nonumber \\ &&-\frac{r^{-6} (\lambda r^{2D} + \kappa q^{2} r^{4})(\lambda r^{2D-4} + \kappa q^2 )^{\frac{2}{D-2}}}{ (\lambda r^{2(D-2)}-\kappa q^2)(D-1)\kappa}.
\end{eqnarray}
The solution looks nasty, but it is not surprising since even in $D=4$ it already appears to be complicated~\cite{Sotani:2014lua, Wei:2014dka}. 

Eqs.~\eqref{36a}, \eqref{37a}, and \eqref{42} constitute a complete static spherically-symmetric solutions of the EiBI-Maxwell system. From the metric part, the solution is asymptotically-flat; {\it i.e.,} $\psi(\infty)$ and $f(\infty)$ (in the absence of $\Lambda$) go to $1$. From the matter part, the EiBI nonlinearity happens to regularize the electric field only in four dimensions. In $D>4$ the electric field still suffers from singularity at the origin, as can be seen from \eqref{37a}. Notice that they are all valid for $\Lambda\geq0$. For AdS, on the other hand, the value $\Lambda=-1/\kappa$ ($\kappa>0$) makes $\lambda=0$. This results in the metric $\psi(r)$ and the electric field $E(r)$ become null. For $\Lambda<-1/\kappa$ ($\lambda<0$) their denominator is not positive-definite (while the $f(r)$'s is), and there is a minimum radius beyond which $\psi(r)$ and $E(r)$ become complex, $r_{min}=\left({\kappa q^2\over|\lambda|}\right)^{1\over2\left(D-2\right)}$. At $r=r_{min}$, both $\psi(r)$ and $E(r)$ blow up. Thus, as in the vacuum case there is a minimum AdS solution in this theory. 

\begin{figure}[htbp]
\centering\leavevmode
\epsfysize=9cm \epsfbox{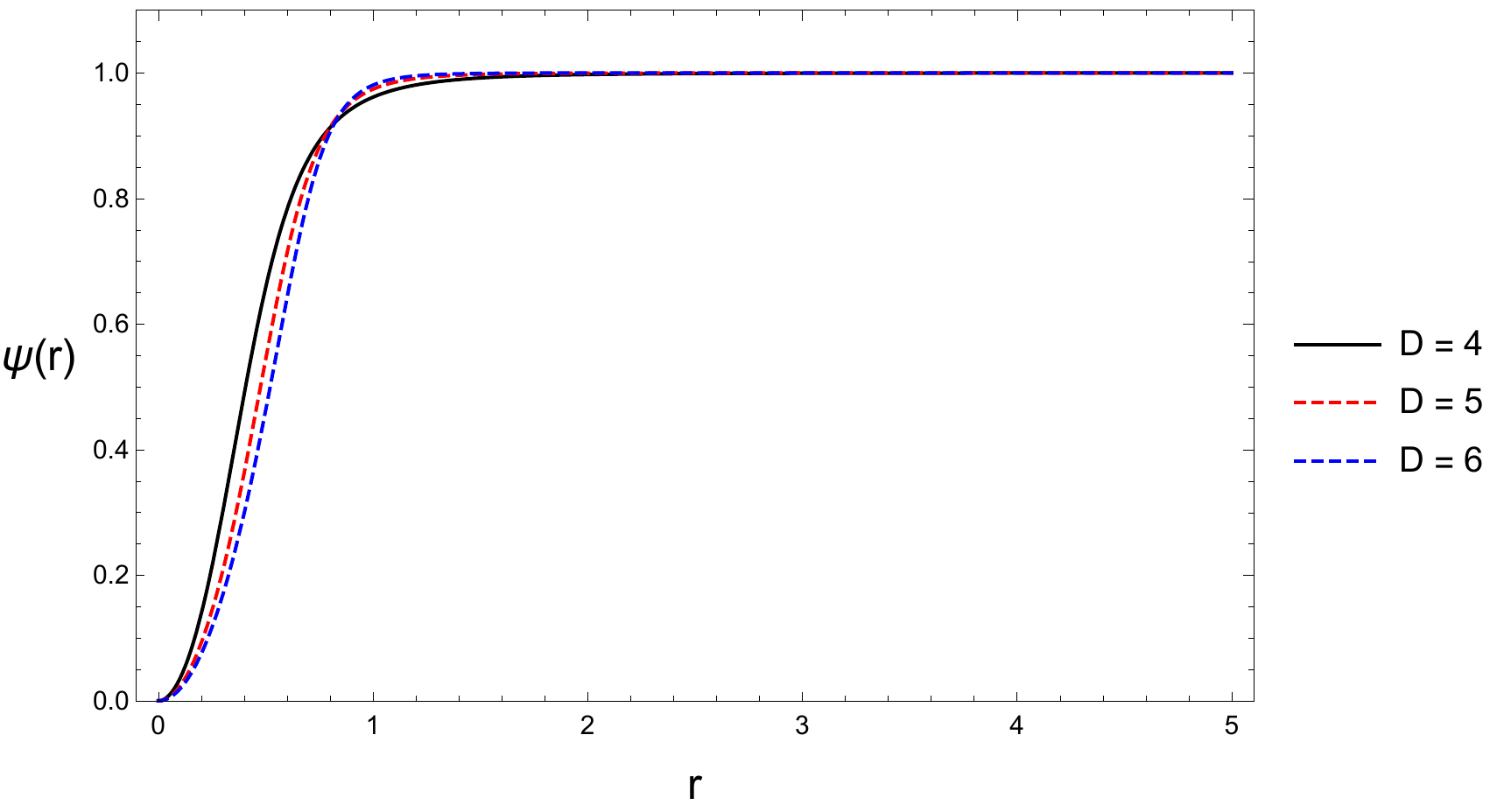}
\caption {A typical plot of $\psi(r)$ with $q=0.5$, $ \kappa=1$ and $\Lambda=0.1$. Changing the parameter values does not significantly alter the shape.}
\label{fig:psidsfig}
\end{figure}

\begin{figure}[htbp]
\centering\leavevmode
\epsfysize=7.5cm \epsfbox{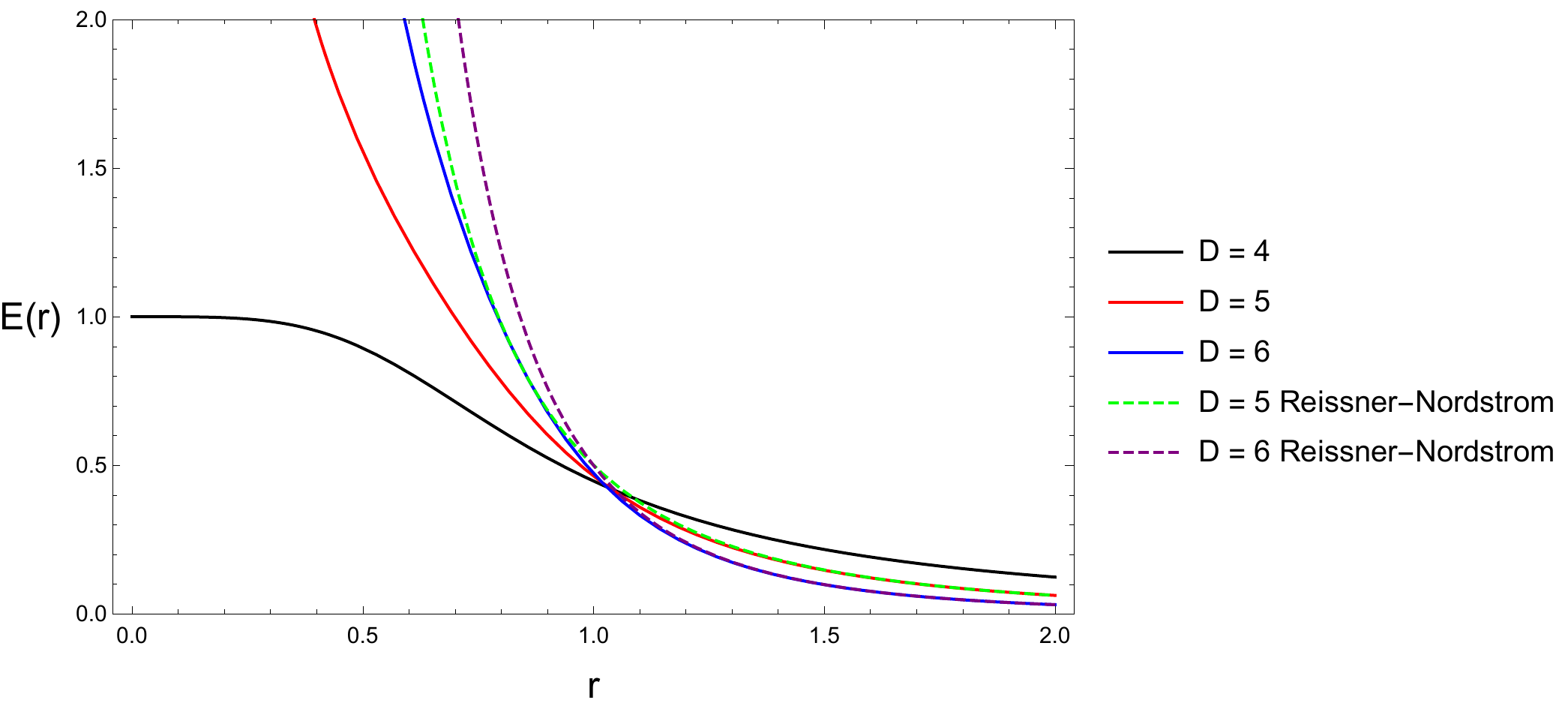}
\caption {A typical plot of $E(r)$ with $ \kappa=1 $, $ q=0.5 $ and $ \Lambda=0 $. The electric field is regular at the origin only in $D=4$.}
\label{medan0}
\end{figure}

In Fig.~\ref{fig:psidsfig} we show the profile of $\psi(r)$ for several $D$. The function interpolates between $0$ and $1$, and the thickness increases as the $D$ gets larger. We plot them for some particular values of $\kappa$ and $q$ in dS space, but the qualitative shape does not significantly differ in asymptotically-flat or AdS spaces, or with different constant parameters. In Fig.~\ref{medan0} we plot a typical profile for the electric field $E(r)$. It is shown that for $D>4$ the field still diverges at the origin.

In order to see whether it describes a black hole or not, we should check if there is singularity enclosed by horizon(s). Not surprisingly the location of singularity coincides with $r_{min}$; {\it i.e.,} when $\lambda>0$, $\psi(r)$ is regular but $f(r)$ blows up, while when $\lambda<0$ the metric $f(r)$ is regular but $\psi(r)$ blows up. Either way, at $r=r_{min}$ the metric becomes singular. Of course to ensure whether at that location the space-time is behaving badly we must calculate the physical invariants; {\it e.g.,} the Ricci scalar constructed from the physical metric, \textcolor{red}{$\mathcal{R}[g]$}. While the general form of such scalar is unilluminating to show, we shall later present its explicit functions in some specific dimensions. 

This blow-up at some non-zero radius value tells us that such singularity is not point-like, but rather surface-like. The next question is whether the radius of event horizon ($r_h$) is greater or smaller than this surface singularity. If $r_h>r_{min}$ we have black hole solutions. On the other hand, if $r_h<r_{min}$ then the singularity is naked. If the solution describes a black hole, then the constant $M$ can be identified as its corresponding mass. Its event horizon(s) can be determined by solving the root(s) of $f(r_h)=0$. If there is more than one horizons, then its extremal state has an extremal radius $r_e$ which satisfies (see, for example, \cite{Myung:2008eb})
\begin{equation}
\label{re1}
f\left(r_e\right)=0,
\end{equation}
and 
\begin{equation}
\label{re2}
f'\left(r_{e}\right)=0.
\end{equation}
The extremal mass can then be expressed as a function of $r_e$, $M(r_e)$. It turns out that $f(r)$ has two horizons for $\Lambda>0$ and only one otherwise ($\Lambda\leq0$). This is rather Schwarzschild-like than Reissner-Nordstrom-like.


\subsection{$D=4$}
In this dimension, our solutions reduce to that obtained in~\cite{Sotani:2014lua, Wei:2014dka}:
\begin{equation}
\label{43}
\psi(r)= \frac{r^2 }{\sqrt{\lambda r^4 + \kappa q^2/\lambda}},
\end{equation}
\begin{equation}
\label{44}
E(r)= \frac{q}{\sqrt{\lambda r^4 + \kappa q^2/\lambda}},
\end{equation}
and
\begin{eqnarray}
\label{46}
f(r)&=&  \frac{r\sqrt{\lambda r^4 + \kappa q^2}}{\lambda r^4 - \kappa q^2} \bigg[ -2\sqrt{\lambda} M + \frac{\left(-\kappa q^2-(\lambda-1)r^4+3\kappa r^2\right) \sqrt{\lambda r^4+\kappa  q^2}}{3 \kappa r^3} \nonumber \\ &&  + \frac{4 i q^2 (\Lambda \kappa -\lambda) }{3 \sqrt{iq\sqrt{\lambda\kappa}}}  \mathcal{F}\left(\left.i \sinh^{-1}\left(r \sqrt{\frac{i \sqrt{\lambda }}{q \sqrt{\kappa }}}\right)\right.,-1\right)\bigg]. 
\end{eqnarray}

\begin{figure}[htbp]
	\centering\leavevmode
	\epsfysize=7.5cm \epsfbox{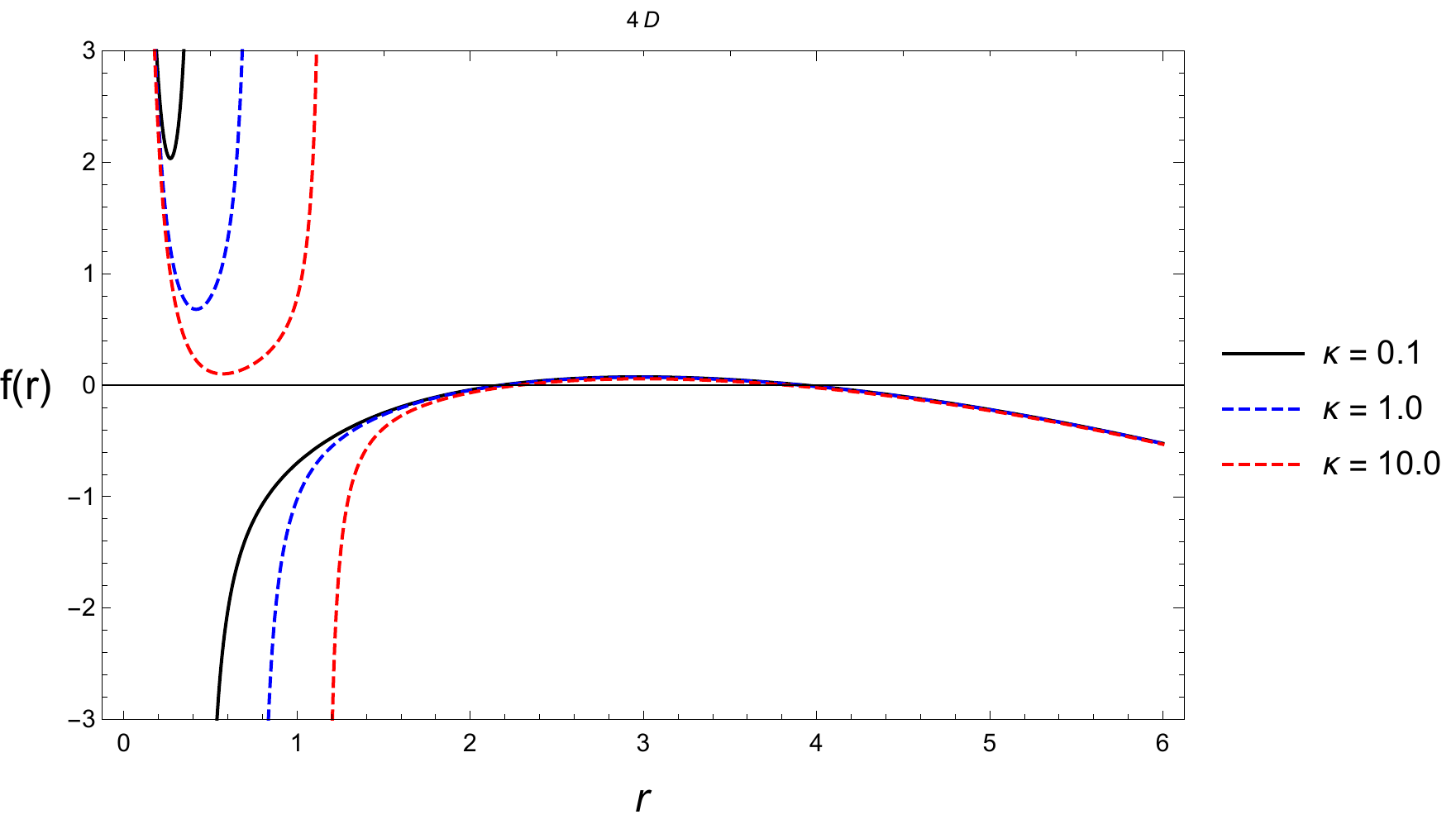}
	\caption {A typical plot of $f(r)$ in $4d$ for various $\kappa$ with $M=1$, $q=0.6$, and $\Lambda=0.1$. The exact numerical values of horizon radii $r_h$ are shown in Table\ref{table:2}.}
	\label{fig:maxwell4ds2}
\end{figure}

The Ricci scalar (for $\lambda=1$) is
\begin{eqnarray}
\mathcal{R} &=& \frac{16 \kappa ^2 q^4}{3 \sqrt{\frac{i}{\sqrt{\kappa } q}} \left(\kappa  q^2-r^4\right)^3 \left(\kappa  q^2 r+r^5\right)^2}
 \bigg[\sqrt{\frac{i}{\sqrt{\kappa } q}} \bigg(3 r^9 \left(3 r^3-7 M \sqrt{\kappa  q^2+r^4}\right) \nonumber \\ && +q^2 r^5 \left(-18 \kappa  M \sqrt{\kappa  q^2+r^4}-2
	r^5+21 \kappa  r^3\right)+\kappa  q^4 r \left(-9 \kappa  M \sqrt{\kappa  q^2+r^4}-8 r^5+15 \kappa  r^3\right) \nonumber \\ && +3 \kappa ^2 q^6 \left(\kappa -2 r^2\right)\bigg)\nonumber\\ &&+2 i q^2 r
	\sqrt{\frac{r^4}{\kappa  q^2}+1} \left(3 \kappa ^2 q^4+6 \kappa  q^2 r^4+7 r^8\right)\ \mathcal{F}\left(\left.i \sinh^{-1}\left(r \sqrt{\frac{i \sqrt{\lambda }}{q \sqrt{\kappa }}}\right)\right.,-1\right) \bigg]. \nonumber \\ 
\end{eqnarray}
It is clear that, as in~\cite{Sotani:2014lua, Wei:2014dka}, the scalar diverges at $r=r_{min}\bigg|_{D=4}$.
\begin{figure}[htbp]
\centering \leavevmode
\epsfysize=7.5cm \epsfbox{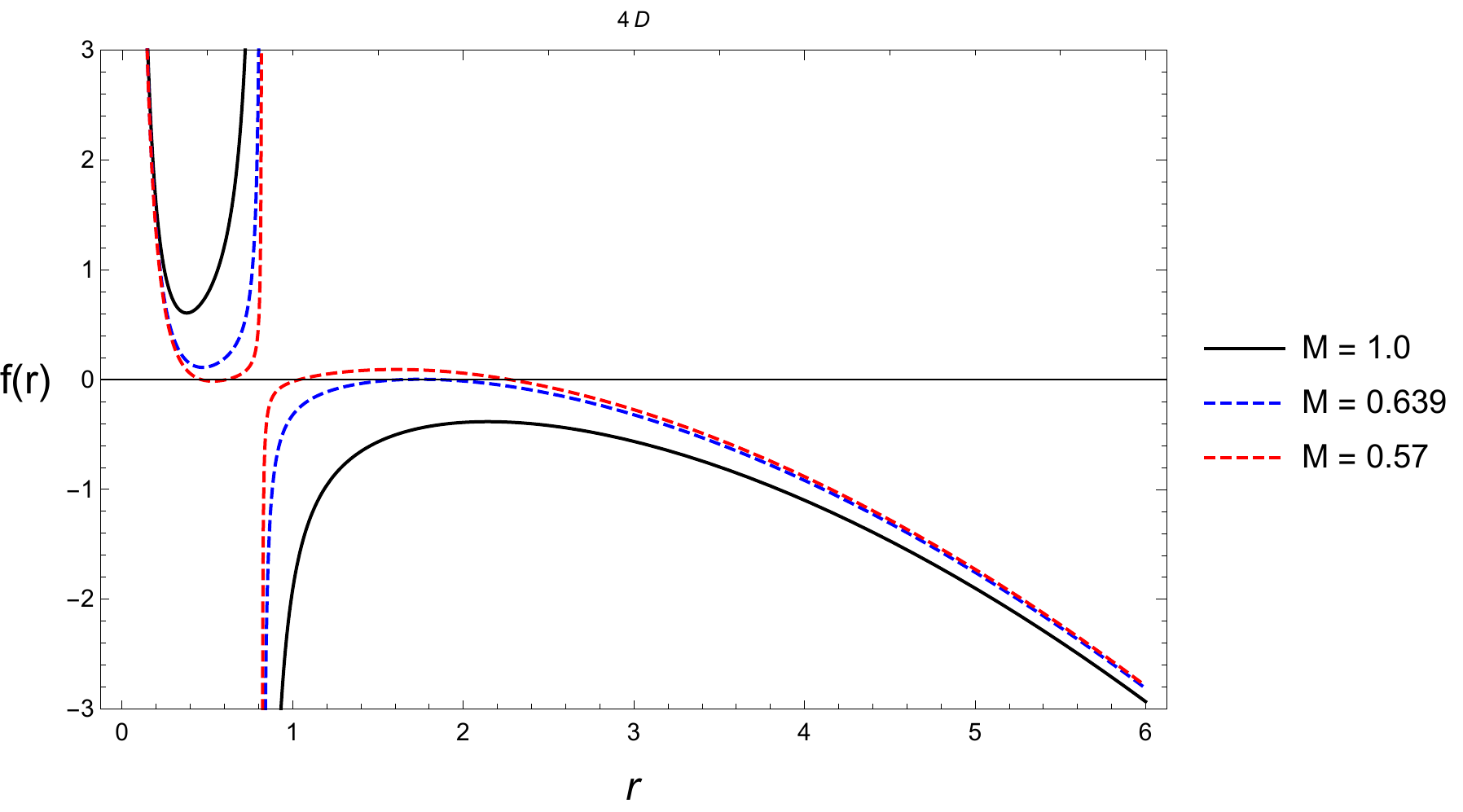}
\caption{A typical plot of $f(r)$ in $4d$ for various $M$ with $\kappa=4$, $q=0.5$, and $\Lambda=0.3$. The blue dashed-line indicates the extremal $ (r_{e}) $ dS black hole. The numerical values of horizon radii $r_h$ is shown in Table~\ref{table:1}. For $\Lambda=0$ the solutions have extensively been plotted in~\cite{Sotani:2014lua, Wei:2014dka}.}
\label{fig:maxwell4ds}
\end{figure}
From the formalism~\eqref{re1}-\eqref{re2} the extremal horizon is determined by solving the following equation
\begin{equation}
-\frac{q^2}{r^2_e} - \Lambda r_e + \frac{1}{r_e}=0.
\end{equation}
The actual expression for $r_e$ is not illuminating analytically, but can be seen numerically in Fig.~\ref{fig:maxwell4ds}. In Fig.~\ref{fig:maxwell4ds2} we show a typical solution in $4d$ with varying $\kappa$. It is shown that $\kappa$ affects the horizon(s), where greater $\kappa$ shifts the horizon(s) to closer radius. Asymptotically, the metric approaches GR's.
\begin{table}[h!]
	\centering
	\begin{tabular}{||c c c c ||} 
		\hline
		$ M $ & $ r_{h1} $& $ r_{e} $  & $ r_{h2} $  \\ [0.5ex] 
		\hline\hline
		1 & -& - & - \\ 
		0.639 & - & 1.68 & - \\
		0.57 & 1.04 & - & 2.21 \\ [1ex] 
		\hline
	\end{tabular}
	\caption{An exact horizon radius $ r_{h} $ from Fig.~\ref{fig:maxwell4ds} with $ r_{min} $ = 0.821.}
	\label{table:1}
\end{table}


\begin{table}[h!]
	\centering
	\begin{tabular}{||c c c c ||} 
		\hline
		$ \kappa $ & $ r_{h1} $ & $ r_{h2} $ & $ r_{min} $ \\ [0.5ex] 
		\hline\hline
		0.1 & 2.18 & 3.92 & 0.43 \\ 
		1.0 & 2.2 & 3.91 & 0.75 \\
		10.0 & 2.29 & 3.82 & 1.15 \\ [1ex] 
		\hline
	\end{tabular}
	\caption{An exact horizon radius $ r_{h} $ and $ r_{min} $ from Fig.~\ref{fig:maxwell4ds2}.}
	\label{table:2}
\end{table}
\subsection{$D=5$}

In this dimension, the solutions are
\begin{equation}
\label{47}
\psi(r) = \frac{r^2}{{(r^{6}+\kappa q^2/\lambda)^{\frac{1}{3}}}},
\end{equation}
\begin{equation}
\label{48}
E(r)= \frac{q}{r \left(r^{6}+\kappa q^2/\lambda\right)^{\frac{1}{3}}},
\end{equation}
and
\begin{eqnarray}
\label{49}
f(r)&=& \frac{r^2 \left(\lambda r^6 + \kappa q^2 \right)^{1/3}}{\left(\lambda r^{6}- \kappa q^2\right)} \bigg[-2M \lambda^{2/3} + \frac{\left(\kappa  q^2+\lambda  r^6\right)^{2/3} q^2 \,
	_2F_1\left(1,1;\frac{4}{3};-\frac{q^2 \kappa }{r^6 \lambda }\right)}{2\lambda  r^6}\nonumber \\ && + \frac{\left(\kappa  q^2+\lambda  r^6\right)^{2/3}
	\left(-\left(\kappa  q^2+\lambda  r^6\right)^{2/3}+r^4+4 \kappa 
	r^2\right)}{4\kappa r^4}  \bigg].
\end{eqnarray}
The Ricci scalar is
\begin{eqnarray}
\mathcal{R} &=& \frac{1}{\kappa  r^6 \left(r^6-\kappa  q^2\right)^3
	\left(\kappa  q^2+r^6\right)^2}\nonumber\\ 
	&&\times\bigg[ \kappa  q^2 \left(\kappa  q^2+r^6\right) \bigg[2 \left(-12 \kappa ^4 q^8-42 \kappa ^3 q^6 r^6-23 \kappa ^2 q^4 r^{12}+4 \kappa  q^2 r^{18}+r^{24}\right) \nonumber \\ &&    \,
	_2F_1\left(1,1;\frac{4}{3};-\frac{q^2 \kappa }{r^6}\right)\nonumber\\&&-3 \left(r^6-\kappa  q^2\right) \left(\kappa  q^2+r^6\right) \left(3 \kappa  q^2+r^6\right) \left(5 \kappa 
	q^2+r^6\right) \, _2F_1\left(1,2;\frac{4}{3};-\frac{q^2 \kappa }{r^6}\right)\bigg] \nonumber \\ &&  +\kappa  q^2 r^{24} \left(24 \kappa  M \sqrt[3]{\kappa  q^2+r^6}-8 r^2 \left(\kappa 
	q^2+r^6\right)^{2/3}+5 r^6-8 \kappa  r^4\right) \nonumber \\ && +4 \kappa ^4 q^8 r^6 \left(24 \kappa  M \sqrt[3]{\kappa  q^2+r^6}+11 r^2 \left(\kappa  q^2+r^6\right)^{2/3}-15 r^6-43 \kappa 
	r^4\right) \nonumber \\ && +\kappa ^3 q^6 r^{12} \left(216 \kappa  M \sqrt[3]{\kappa  q^2+r^6}+79 r^2 \left(\kappa  q^2+r^6\right)^{2/3}-65 r^6-252 \kappa  r^4\right) \nonumber \\ && +\kappa ^2 q^4 r^{18}
	\left(240 \kappa  M \sqrt[3]{\kappa  q^2+r^6}+23 r^2 \left(\kappa  q^2+r^6\right)^{2/3}-7 r^6-116 \kappa  r^4\right)\nonumber\\ &&+5 r^{32} \left(\left(\kappa 
	q^2+r^6\right)^{2/3}-r^4\right) +\kappa ^5 q^{10} r^2 \left(\left(\kappa  q^2+r^6\right)^{2/3}-12 r^4-28 \kappa  r^2\right) \bigg].
\end{eqnarray}
It is obvious that we have surface singularity at $r=r_{min}\bigg|_{D=5}=\left(\kappa q^2\right)^{1/6}$. As in the case of $D=4$, the extremal black hole can be found by solving the following polynomial equation
\begin{equation}
\kappa  q^2 \left(\left(\kappa  q^2+\lambda  r^6\right)^{2/3}-2 r^2 \left(\kappa +r^2\right)\right)+\lambda  r^6 \left(-\left(\kappa  q^2+\lambda  r^6\right)^{2/3}+r^4+2 \kappa 
r^2\right) = 0.
\end{equation}

\subsection{$D=6$}

The solutions in $D=6$ are
\begin{equation}
\label{50}
\psi(r) = \frac{r^2}{{(\lambda r^{8}+\kappa q^2/\lambda)^{\frac{1}{4}}}},
\end{equation}
\begin{equation}
\label{51}
E(r)= \frac{q}{r^{2}\left(\lambda r^{8}+\kappa q^2/\lambda\right)^{\frac{1}{4}}},
\end{equation}
and
\begin{eqnarray}
\label{52}
f(r)&=& \frac{r^3 \left(\lambda r^8 + \kappa q^2 \right)^{1/4}}{\left(\lambda r^{8}- \kappa q^2\right)} \bigg[-2M \lambda^{3/4}-\frac{\left(\kappa  q^2+\lambda  r^8\right)^{3/4} 20 \lambda  r^{12} \,
	_2F_1\left(1,\frac{13}{8};\frac{15}{8};-\frac{r^8 \lambda }{q^2 \kappa }\right)}{35 \kappa ^2 q^2 r^5} \nonumber \\ && + \frac{7
	\kappa  q^2  \left(\kappa  q^2+\lambda  r^8\right)^{3/4} \left(\sqrt{\kappa  q^2+\lambda  r^8}-5 r^2 \left(\kappa
	+r^2\right)\right)}{35 \kappa ^2 q^2 r^5} \bigg] .         
\end{eqnarray}

The $6d$ Ricci scalar is
\begin{eqnarray}
\mathcal{R} &=& -\frac{2}{35 \kappa ^2 q^2 r^4 \left(\kappa  q^2-r^8\right)^3 \left(\kappa  q^2+r^8\right)^2}\nonumber\\
&&\times  \bigg[20 r^{12} \left(\kappa  q^2+r^8\right) \bigg[8 \left(r^8-\kappa  q^2\right) \left(\kappa  q^2+r^8\right) \nonumber \\ && \left(-13	\kappa ^2 q^4-6 \kappa  q^2 r^8+3 r^{16}\right)     \, _2F_1\left(\frac{13}{8},2;\frac{15}{8};-\frac{r^8}{q^2 \kappa
	}\right)\nonumber\\&&+\bigg[27 \kappa ^4 q^8+168 \kappa ^3 q^6 r^8+78 \kappa ^2 q^4 r^{16}\nonumber\\&&-32 \kappa  q^2 r^{24}+15 r^{32}\bigg] \,
	_2F_1\left(1,\frac{13}{8};\frac{15}{8};-\frac{r^8}{q^2 \kappa }\right)\bigg]\nonumber\\&&+7 \kappa  q^2 \bigg[\kappa  q^2 r^{29} \bigg[160
	\kappa  M \sqrt[4]{\kappa  q^2+r^8} \nonumber \\ && +19 r^3 \sqrt{\kappa  q^2+r^8}-65 r^7-10 \kappa  r^5\bigg]\nonumber\\&&+\kappa ^4 q^8 r^5 \bigg[400
	\kappa  M \sqrt[4]{\kappa  q^2+r^8}-151 r^3 \sqrt{\kappa  q^2+r^8}+975 r^7+800 \kappa  r^5\bigg]\nonumber\\&&+10 \kappa ^3 q^6 r^{13}
	\left(96 \kappa  M \sqrt[4]{\kappa  q^2+r^8}-27 r^3 \sqrt{\kappa  q^2+r^8}+119 r^7+124 \kappa  r^5\right) \nonumber \\ && +10 \kappa ^2 q^4
	r^{21} \left(104 \kappa  M \sqrt[4]{\kappa  q^2+r^8}-9 r^3 \sqrt{\kappa  q^2+r^8}+23 r^7+42 \kappa  r^5\right)\nonumber\\ &&+15 r^{40}
	\left(-\sqrt{\kappa  q^2+r^8}+5 r^4+4 \kappa  r^2\right)\nonumber\\ &&+5 \kappa ^5 q^{10} \left(-\sqrt{\kappa  q^2+r^8}+31 r^4+10 \kappa 
	r^2\right)\bigg]\bigg],
\end{eqnarray}
which shows singularity at $r=r_{min}\bigg|_{D=6}=\left(\kappa q^2\right)^{1/8}$. The polynomial equation for  the extremal horizon is surprisingly simpler than its $5d$ counterpart, 
\begin{equation}
3r^2 +r^4-\sqrt{\kappa  q^2+\lambda  r^8} \kappa  =0.
\end{equation}



\section{EiBI COUPLED TO B-I ELECTRODYNAMICS IN HIGHER DIMENSIONS}
\label{sec:bieldin}

From the previous section we learn that in higher dimensions ($D>4$) the nonlinearity of EiBI gravity cannot uplift the divergence of charged particle self-energy. It is then tempting to investigate whether regular EM field can exist in nonlinear electrodynamics. While there are numerous models of nonlinear electrodynamics that exist in literature, it is without doubt that the Born-Infeld (B-I)~\cite{Born:1934gh} is one of the most popular one. In this work we limit ourselves to study the exact solutions of higher-dimensional EiBI gravity with non-vanishing cosmological constant coupled to B-I electrodynamics. This theory possesses two coupling parameters that control the strength of nonlinearity of both the gravity ($\kappa$) and electrodynamics ($b$) sectors. The $4d$ counterpart of this theory has been addressed in~\cite{Jana:2015cha}.

The matter Lagrangian density is given by
\begin{equation}
\label{53}
\mathcal{L}_{M}=\frac{b^2}{4\pi}\left[1-\sqrt{1+\frac{F_{\mu\nu}F^{\mu\nu}}{2b^2}}\right],
\end{equation}
where $b$ is a nonlinear parameter which reduces the lagrangian to Maxwell in the weak-coupling limit, $b\rightarrow\infty$. The resulting energy-momentum tensor has the following general expression
\begin{equation}
\label{54}
T_{\mu\nu}= -\frac{b^2}{4\pi}\left[g_{\mu\nu}\left(\sqrt{1+\frac{F_{\alpha\beta}F^{\alpha\beta}}{2b^2}}-1\right)-\frac{F_{\mu\sigma} F_{\nu}^{\sigma}}{b^2 \sqrt{1+\frac{F_{\alpha\beta}F^{\alpha\beta}}{2b^2}}}\right].
\end{equation}
Variation with respect to $A_{\mu}$ yields 
\begin{equation}
\label{55}
\nabla_{\mu}\left(\frac{F^{\mu\nu}}{\sqrt{1+\frac{F_{\alpha\beta}F^{\alpha\beta}}{2b^2}}}\right)=0.
\end{equation}

Here, we note that the metric ansatz~\eqref{5}-\eqref{6} produce complicated equations which cannot be solved easily. We therefore follow~\cite{Jana:2015cha} to assume another form of the metrics,
\begin{equation}
\label{56}
g_{\mu\nu}dx^{\mu}dx^{\nu}=-U(\bar{r})e^{2\psi(\bar{r})}dt^2+U(\bar{r})e^{2\nu(\bar{r})}d\bar{r}^2+V(\bar{r})\bar{r}^2d\Omega^2_{D-2},
\end{equation}
\begin{equation}
\label{57}
q_{\mu\nu}dx^{\mu}dx^{\nu}=-e^{2\psi(\bar{r})}dt^2+e^{2\nu(\bar{r})}d\bar{r}^2+\bar{r}^2d\Omega^2_{D-2},
\end{equation}
where $U, V, \psi,$ and $\nu$ are the functions to be solved along with the electric field $E(r)$. 

It is not difficult to solve the electromagnetic field equation to obtain
\begin{equation}
\label{58}
E(\bar{r})=\frac{qUe^{\nu+\psi}}{\sqrt{V^{D-2}\bar{r}^{2(D-2)}+\frac{q^2}{b^2}}}
\end{equation}
with $q$, as before, is the electric charge.

The non-zero energy-momentum tensor components are
\begin{equation}
\label{59}
T^{00}=\frac{e^{-2\psi}b^2}{U 4\pi}\left(\frac{\sqrt{V^{D-2}\bar{r}^{2(D-2)}+\frac{q^{2}}{b^{2}}}}{V^{\frac{D-2}{2}}\bar{r}^{D-2}}-1 \right),
\end{equation}
\begin{equation}
\label{60}
T^{11}=-\frac{e^{-2\nu}b^2}{U 4\pi}\left(\frac{\sqrt{V^{D-2}\bar{r}^{2(D-2)}+\frac{q^{2}}{b^{2}}}}{V^{\frac{D-2}{2}}\bar{r}^{D-2}}-1 \right),
\end{equation}
\begin{equation}
\label{61}
T^{22}=\frac{b^2}{ 4\pi V\bar{r}^{2}}\left(1-\frac{V^{\frac{D-2}{2}}\bar{r}^{D-2}}{\sqrt{V^{D-2}\bar{r}^{2(D-2)}+\frac{q^{2}}{b^{2}}}} \right),
\end{equation}
\begin{equation}
T^{33}= \frac{T^{22}}{\sin^2\theta},\ \ \ \ T^{ab}=\frac{T^{22}}{\prod_{j=1}^{D-2}\sin^2\theta_j}.
\end{equation}

The $00$- (or $11$-) component of eqs.~\eqref{4} yields the following polynomial equation for $V$
\begin{equation}
\label{115}
\lambda^2\left(1-\tilde{\alpha}\right)V^{D-2}-\lambda\left(2-\tilde{\alpha}\right)V^{\frac{D-2}{2}}+1-\frac{\kappa q^{2}\lambda\tilde{\alpha}}{\bar{r}^{2(D-2)}}= 0,
\end{equation}
where\footnote{The case of $\alpha=4\kappa b^2$ has been defined in~\cite{Jana:2015cha}.} $\tilde{\alpha}\equiv4\kappa b^2/\lambda$. From the $22$-component we can get $U$,
\begin{equation}
\label{115a}
U(\bar{r})=\frac{\left(1-\frac{\tilde{\alpha}}{2}\right)\sqrt{1+\frac{4 \kappa \lambda q^2}{ \tilde{\alpha} \bar{r}^{2(D-2)}} -\frac{4\kappa q^2 \lambda}{\bar{r}^{2(D-2)}}     } - \frac{\tilde{\alpha}}{2}  }{\left(\lambda-\tilde{\alpha} \lambda \right) \sqrt{1+\frac{4 \kappa \lambda q^2}{ \tilde{\alpha} \bar{r}^{2(D-2)}} -\frac{4\kappa q^2 \lambda}{\bar{r}^{2(D-2)}}     } }~ V(\bar{r})^{\frac{4-D}{2}}
\end{equation}
On the other hand, from \eqref{2} one obtains
\begin{equation}
\label{65}
\left(\frac{1-U}{\kappa}\right) e^{2\psi}=e^{-2\nu+2\psi}\left(-\frac{(D-2)\psi'}{\bar{r}}+\nu'\psi'-\psi'^{2}-\psi'' \right),
\end{equation}
\begin{equation}
\label{66}
\left(\frac{1-U}{\kappa}\right) e^{2\nu}=\left(\frac{(D-2)\nu'}{\bar{r}}+\nu'\psi'-\psi'^{2}-\psi'' \right),
\end{equation}
\begin{equation}
\label{67}
\left(\frac{1-V}{\kappa}\right)\bar{r}^{2}= (D-3)-\frac{1}{(D-2)\bar{r}^{D-4}}\left(\left(\bar{r}^{D-2}\right)' e^{2\psi}\right)'.
\end{equation}
The first two are solved by $\psi=-\nu$. Inserting it into \eqref{67} yields
\begin{equation}
\label{69}
e^{2\psi(\bar{r})}=1-\frac{2M}{\bar{r}^{(D-3)}}-\frac{\bar{r}^{2}}{(D-1)\kappa}+\frac{1}{\kappa \bar{r}^{(D-3)}} \int V(\bar{r})\bar{r}^{(D-2)} d\bar{r}.
\end{equation}
Eqs.~\eqref{115}, \eqref{115a}, and \eqref{69} constitute a set of system of equations needed to obtain the metric solutions. Unlike its $4d$ counterpart~\cite{Jana:2015cha}, Eq.~\eqref{69} cannot be integrated analytically for any general solution of~\eqref{115}. We therefore have to resort to consider some specific values of $\tilde{\alpha}$ that makes the integration solvable.

\subsection{$\tilde{\alpha}=1$}

We observe that for $\tilde{\alpha}=1$ the metric $e^{2\psi}$ can be integrated exactly. Eq.~\eqref{115} greatly reduces to a linear equation in $V^{D-2\over2}$,
\begin{equation}
\label{116}
-\frac{1}{\lambda} V^{\frac{D-2}{2}} +\frac{1}{\lambda} - \frac{\kappa q^2}{\bar{r}^{2(D-2)}} = 0.
\end{equation}
It is then trivial to show
\begin{equation}
\label{117}
V(\bar{r})= \left(\frac{1}{\lambda}-\frac{\kappa q^2}{\bar{r}^{2(D-2)}}\right)^{\frac{2}{D-2}},
\end{equation}
\begin{equation}
\label{118}
U(\bar{r})=\left(\frac{1}{\lambda}+\frac{\kappa q^2}{\bar{r}^{2(D-2)}} \right) V(\bar{r})^{\frac{4-D}{2}},
\end{equation}
and
\begin{eqnarray}
\label{119}
e^{2\psi}(\bar{r})&=& 1-\frac{2M}{\bar{r}^{(D-3)}}-\frac{\bar{r}^{2}}{(D-1)\kappa}  + \frac{\bar{r}^2 \,
	_2F_1\left(-\frac{2}{D-2},-\frac{D-1}{2 (D-2)};\frac{D-3}{2 (D-2)};q^2 \bar{r}^{4-2 D} \kappa  \lambda \right)}{\lambda^{\frac{2}{D-2}}(D-1) \kappa}. 
\end{eqnarray}
Eqs.~\eqref{117}-\eqref{119} constitute a set of exact solutions of EiBI-BI theory in higher dimensions.

\begin{figure}[htbp]
	\centering\leavevmode
	\epsfysize=7.5cm \epsfbox{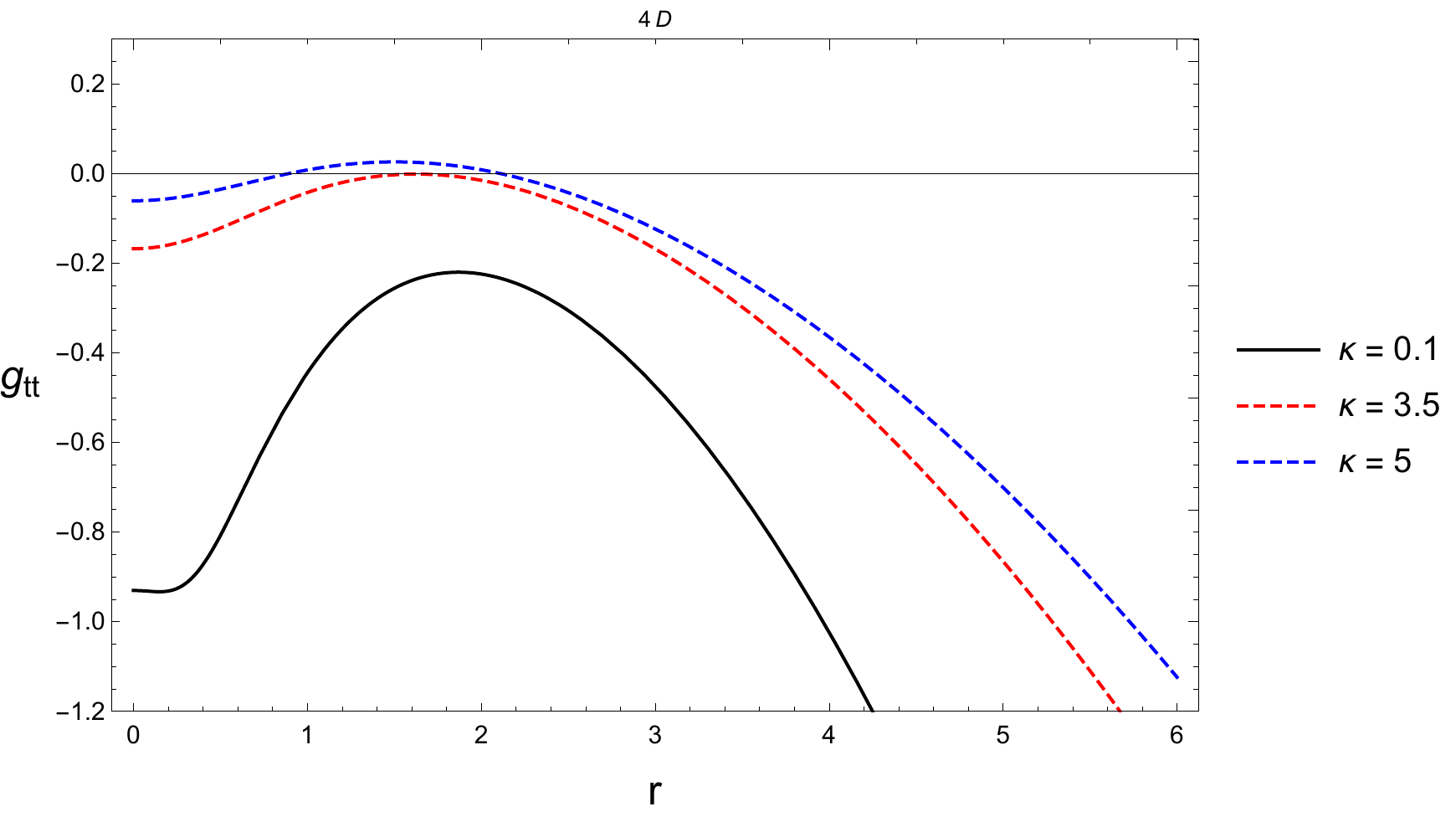}
	\caption {A typical plot of $4d$ $ g_{tt} $ with $ M=1$, $ q=0.8 $ and $ \Lambda=0.3$.}
	\label{fig:gtt4dskappa}
\end{figure}
\begin{table}[h!]
	\centering
	\begin{tabular}{||c c c ||} 
		\hline
		$ \kappa $ & $ r_{h1} $& $ r_{h2} $ \\ [0.5ex] 
		\hline\hline
		0.1 & -& - \\ 
		3.5 & - & - \\
		5.0 & 0.89 & 2.1 \\ [1ex] 
		\hline
	\end{tabular}
	\caption{An exact horizon radius $ r_{h} $ from Fig.~\ref{fig:gtt4dskappa}.}
	\label{table:3}
\end{table}

This solution is found in the gauge~\eqref{56}, different from the Tangherlini's. Fortunately it is possible to bring it into the gauge~\eqref{5} by setting $r^2 = \left(\frac{1}{\lambda}-\frac{\kappa q^2}{\bar{r}^{2(D-2)}}\right)^{\frac{2}{D-2}} \bar{r}^2 $. The resulting metric is
\begin{equation}
\label{122}
ds^2 = - A(r) f(r) dt^2 + \frac{ r^{D-4} \left(\frac{\lambda}{2}\right)^{\frac{2}{D-2}} A(r) \left(1+\sqrt{B(r)_D}\right)^{\frac{2}{D-2}}}{ B(r)_D f(r)} dr^2 + r^2 d\Omega^2
\end{equation}
where
\begin{equation}
\label{122a}
A(r) = \left(\frac{1}{\lambda}+\frac{\kappa q^2}{y(r)^{2(D-2)}} \right) \left(\frac{1}{\lambda}- \frac{\kappa q^2}{y(r)^{2(D-2)}}\right)^{\frac{4-D}{D-2}},
\end{equation}
\begin{equation}
\label{123}
f(r)= 1-\frac{2M}{y(r)^{(D-3)}}-\frac{y(r)^{2}}{(D-1)\kappa}  + \frac{y(r)^2 \,_2F_1\left(-\frac{2}{D-2},-\frac{D-1}{2 (D-2)};\frac{D-3}{2 (D-2)};{\kappa q^2 \lambda\over y(r)^{2(D-2)}} \right)}{\lambda^{\frac{2}{D-2}}(D-1) \kappa },
\end{equation}

\begin{figure}[htbp]
	\centering\leavevmode
	\epsfysize=7.5cm \epsfbox{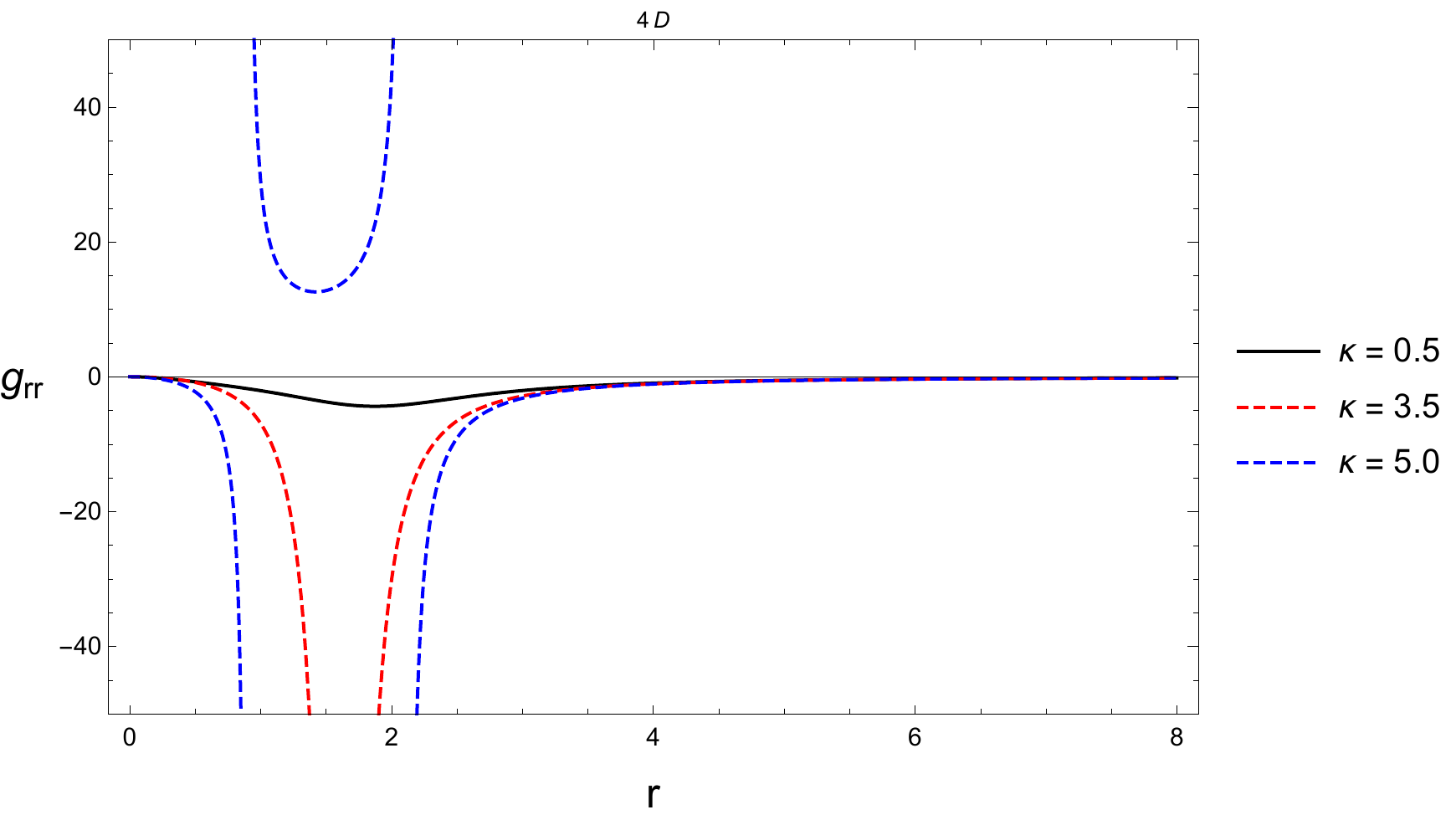}
	\caption {A typical plot of $4d$ $ g_{rr} $ with $ M=1$, $ q=0.8 $ and $ \Lambda=0.3$.}
	\label{fig:grr4dskappa}
\end{figure}

\begin{equation}
\label{124}
y(r)\equiv\bar{r}(r)=\left({\lambda \over 2}\right)^{\frac{1}{D-2}} \left(1+\sqrt{B(r)_D}\right)^{\frac{1}{D-2}}\ r,
\end{equation}
and
\begin{equation}
\label{122b}
B(r)_{D}= 1+ \frac{4 \kappa q^2}{\lambda r^{2(D-2)}}.
\end{equation}

\subsubsection{$D=4$}

In $4d$ our general solutions reduce to that of Jana-Kar~\cite{Jana:2015cha},
\begin{equation}
\label{109}
V(\bar{r})=\frac{1}{\lambda }-\frac{\kappa  q^2}{\bar{r}^4},
\end{equation}
\begin{equation}
\label{110}
U(\bar{r})= \left(\frac{1}{\lambda}+\frac{\kappa q^2}{\bar{r}^{4}} \right)
\end{equation}
and
\begin{equation}
\label{111}
e^{2\psi(\bar{r})}=1-\frac{2M}{\bar{r}}-\left(\frac{\lambda-1}{3\lambda\kappa}\right)\bar{r}^2 + \frac{q^2}{\bar{r}^2}.
\end{equation}

\begin{figure}[htbp]
	\centering\leavevmode
	\epsfysize=7.5cm \epsfbox{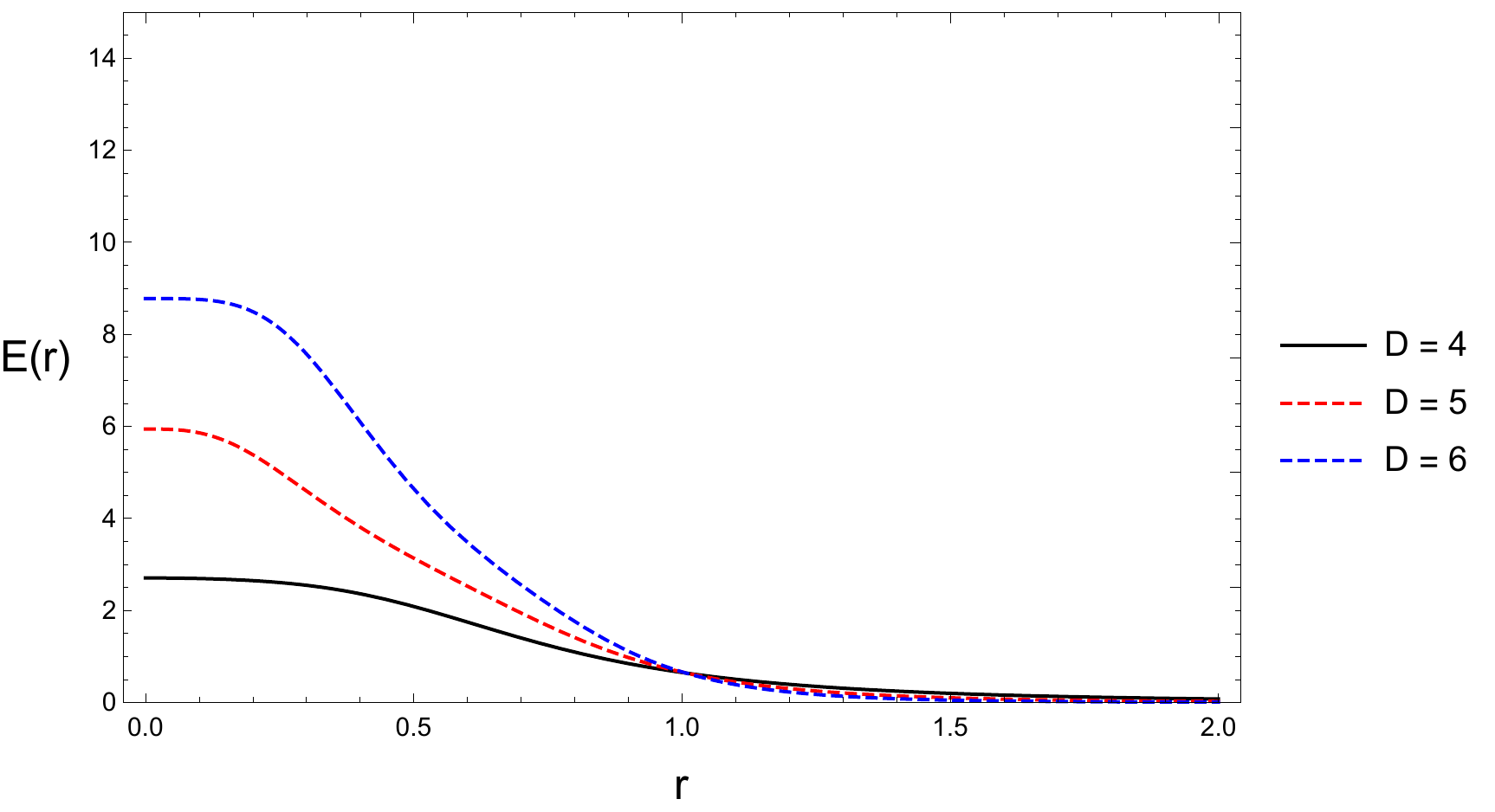}
	\caption {Plot $ E(r) $ with $ \kappa=0.1 $, $ q=0.7 $ and $ b=1.5 $ in asymptotically-flat space-time.}
	\label{weakweak}
\end{figure}
In the Tangherlini gauge, $r^{2}=  \left(\frac{1}{\lambda }-\frac{\kappa  q^2}{r^4}\right)\bar{r}^2$, the metric reads
\begin{equation}
\label{113}
ds^2 = -A(r) f(r) dt^2 + \frac{\left(\lambda \over 2\right) A(r) \left(1+ \sqrt{B(r)_4}\right)}{ B(r)_4 f(r)} dr^2 + r^2 d\Omega^2,
\end{equation}
where 
\begin{equation}
A(r)=  \left(\frac{1}{\lambda }+\frac{4 \kappa  q^2}{\left(\sqrt{\lambda } \sqrt{4 \kappa  q^2+\lambda  r^4}+\lambda 
	r^2\right)^2}\right),
\end{equation}
\begin{eqnarray}
\label{114}
f(r)&=&\bigg[1-\frac{2 \sqrt{2} M}{\sqrt{\sqrt{\lambda } \sqrt{4 \kappa  q^2+\lambda r^4}+\lambda  r^2}} + \frac{2 q^2}{\sqrt{\lambda } \sqrt{4 \kappa  q^2+\lambda  r^4}+\lambda  r^2}  \nonumber \\ && - \frac{\left(\lambda-1\right)r^2}{6\kappa} \left(\frac{\sqrt{4 \kappa  q^2+\lambda r^4}}{\sqrt{\lambda } r^2}+1\right)\bigg],
\end{eqnarray}
and
\begin{equation}
B(r)_{4}= 1+ \frac{4 \kappa q^2}{\lambda r^{4}}.
\end{equation}
Profiles of metric solutions in the Tangherlini gauge can be seen in Figs.~\ref{fig:gtt4dskappa}-\ref{fig:grr4dskappa}. Note that all solutions we found are regular at the origin\footnote{This, of course, cannot be taken as a sign of regular black hole (as in the case of Bardeen black hole~\cite{AyonBeato:1998ub}.) The scalar curvature in Fig.~\ref{fig:ricci} shows that all solutions in all dimensions develop singularity at the origin. Thus, the regularity of black hole solutions indicate that the electromagnetic source is point-charge.}, as were also shown in~\cite{Jana:2015cha}. It is interesting that this typical point-charge solution only appears in four dimensions. In higher dimensions, as can be seen later, all metric solutions are singular at the origin.

\begin{figure}[htbp]
	\centering\leavevmode
	\epsfysize=7.5cm \epsfbox{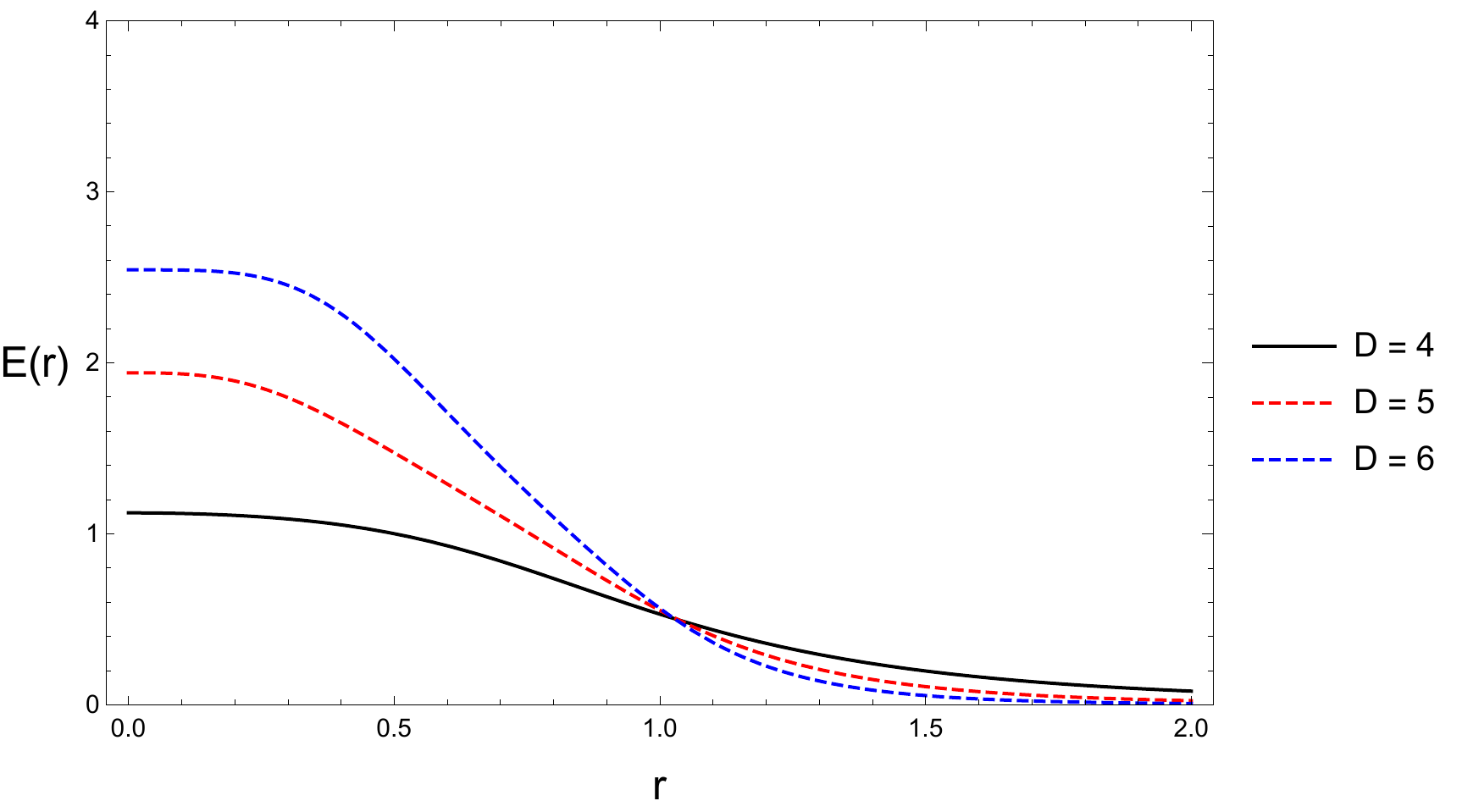}
	\caption {Plot $ E(r) $ with $ \kappa=0.4 $, $ q=0.7 $ and $ b=0.7 $ in asymptotically-flat space-time.}
	\label{weakstrong}
\end{figure}
\begin{figure}[htbp]
	\centering\leavevmode
	\epsfysize=7.5cm \epsfbox{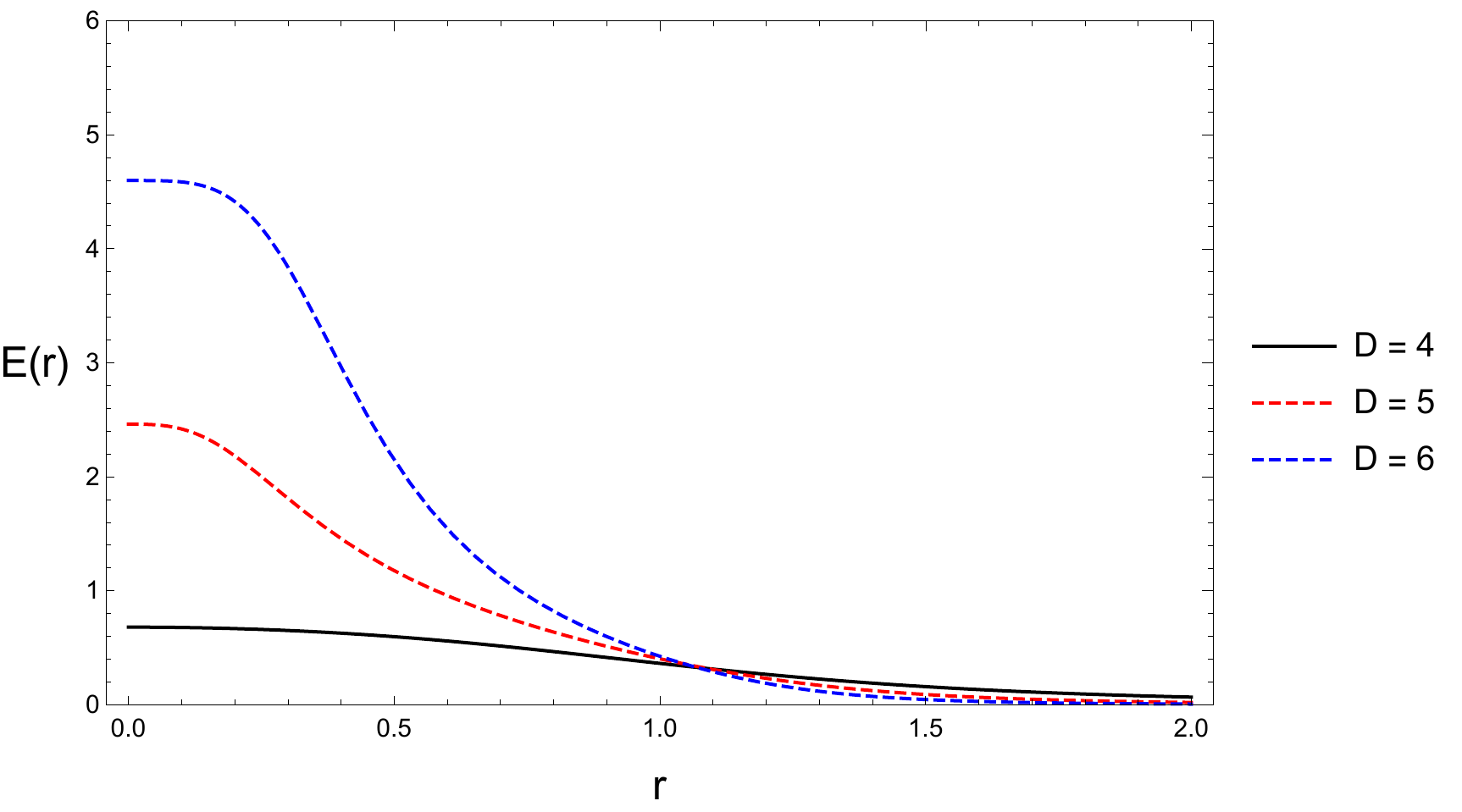}
	\caption {Plot $ E(r) $ with $ \kappa=2.0 $, $ q=0.7 $ and $ b=0.35 $ in asymptotically-flat space-time.}
	\label{strongstrong}
\end{figure}

The solutions for electric fields can be inferred from~\eqref{58} by inserting~\eqref{117} and \eqref{118}. Typical solutions for electric fields can be seen from Figs.~\ref{weakweak}-\ref{strongstrong}. Here the three profiles represent the three possible family of solutions characterized by $\kappa$ and $b$, respectively. They are: (i) a weak EiBI ($0<\kappa<1$) and (marginally-)weak BI ($b^2\gtrsim1$) in Fig.~\ref{weakweak}, (ii) a weak EiBI ($0<\kappa<1$) and strong BI ($b^2<1$) in Fig.~\ref{weakstrong}, and (iii) a (marginally-)strong EiBI ($\kappa\gtrsim1$) and strong BI ($b^2<1$) in Fig.~\ref{strongstrong}. The limiting value of case (i) is the (marginally) weakly-charged RN-(A)dS, while (ii) is the Einstein-BI~\cite{Dey:2004yt, Cai:2004eh}. That some of them are only {\it marginally} weak/strong comes from the fact that we can only solve them in the condition where $\tilde{\alpha}=1$; {\it i.e.,} there is a constraint that $\kappa$ and $b$ have to satisfy. Precisely due to this constraint that we cannot unfortunately obtain solutions in the limiting value of the previously-solved EiBI-Maxwell case; {\it i.e.,} the value $\tilde{\alpha}=1$ does not allow us to have strong $\kappa$ and weak $b^2$ simultaneously.

\subsubsection{$D=5$}

In dimensions higher than four, the metric solutions cannot be expressed in terms of polynomial (or closed) functions,
\begin{equation}
\label{125}
V(\bar{r})= \left(\frac{1}{\lambda }-\frac{\kappa  q^2}{\bar{r}^6}\right)^{2/3},
\end{equation}
\begin{equation}
\label{126}
U(\bar{r}) =  \left(\frac{1}{\lambda}+\frac{\kappa q^2}{\bar{r}^{6}} \right) V(\bar{r})^{-\frac{1}{2}},
\end{equation}
and
\begin{eqnarray}
\label{127}
e^{2\psi(\bar{r})}&=& 1-\frac{2M}{\bar{r}^{2}}-\frac{\bar{r}^{2}}{4\kappa}+ \frac{\bar{r}^2 \, _2F_1\left(-\frac{2}{3},-\frac{2}{3};\frac{1}{3};\frac{q^2 \kappa \lambda
	}{\bar{r}^6}\right)}{4 \kappa  \lambda ^{2/3}}. \nonumber \\&&
\end{eqnarray}

Transforming the radial coordinate $r^2\rightarrow\left(\frac{1}{\lambda }-\frac{\kappa  q^2}{\bar{r}^6}\right)^{2/3} \bar{r}^2 $ we get
\begin{equation}
\label{128}
ds^2 = - A(r) f(r) dt^2 + \frac{ r \left(\frac{\lambda}{2}\right)^{\frac{2}{3}} A(r) \left(1+\sqrt{B(r)_5}\right)^{\frac{2}{3}}}{ B(r)_5 f(r)} dr^2 + r^2 d\Omega^2,
\end{equation}

where
\begin{equation}
A(r)=\frac{ \frac{1}{\lambda} + \frac{4\kappa q^2}{\lambda^2 r^6 \left(1+\sqrt{B(r)_5}\right)^2}  }{  \left(\frac{1}{\lambda} - \frac{4\kappa q^2}{\lambda^2 r^6 \left(1+\sqrt{B(r)_5}\right)^2}     \right)^{\frac{1}{3}}},
\end{equation}
\begin{eqnarray}
\label{129}
f(r)= 1-\frac{2M}{y(r)^{2}}-\frac{y(r)^{2}}{4\kappa}+ \frac{y(r)^2 \, _2F_1\left(-\frac{2}{3},-\frac{2}{3};\frac{1}{3};\frac{q^2 \kappa \lambda	}{y(r)^6}\right)}{4 \kappa  \lambda ^{2/3}}, 
\end{eqnarray}
\begin{equation}
y(r)\equiv\bar{r}(r)=\left({\lambda \over 2}\right)^{\frac{1}{3}} \left(1+\sqrt{B(r)_5}\right)^{\frac{1}{3}}\ r,
\end{equation}
and

\begin{figure}[htbp]
	\centering\leavevmode
	\epsfysize=7.5cm \epsfbox{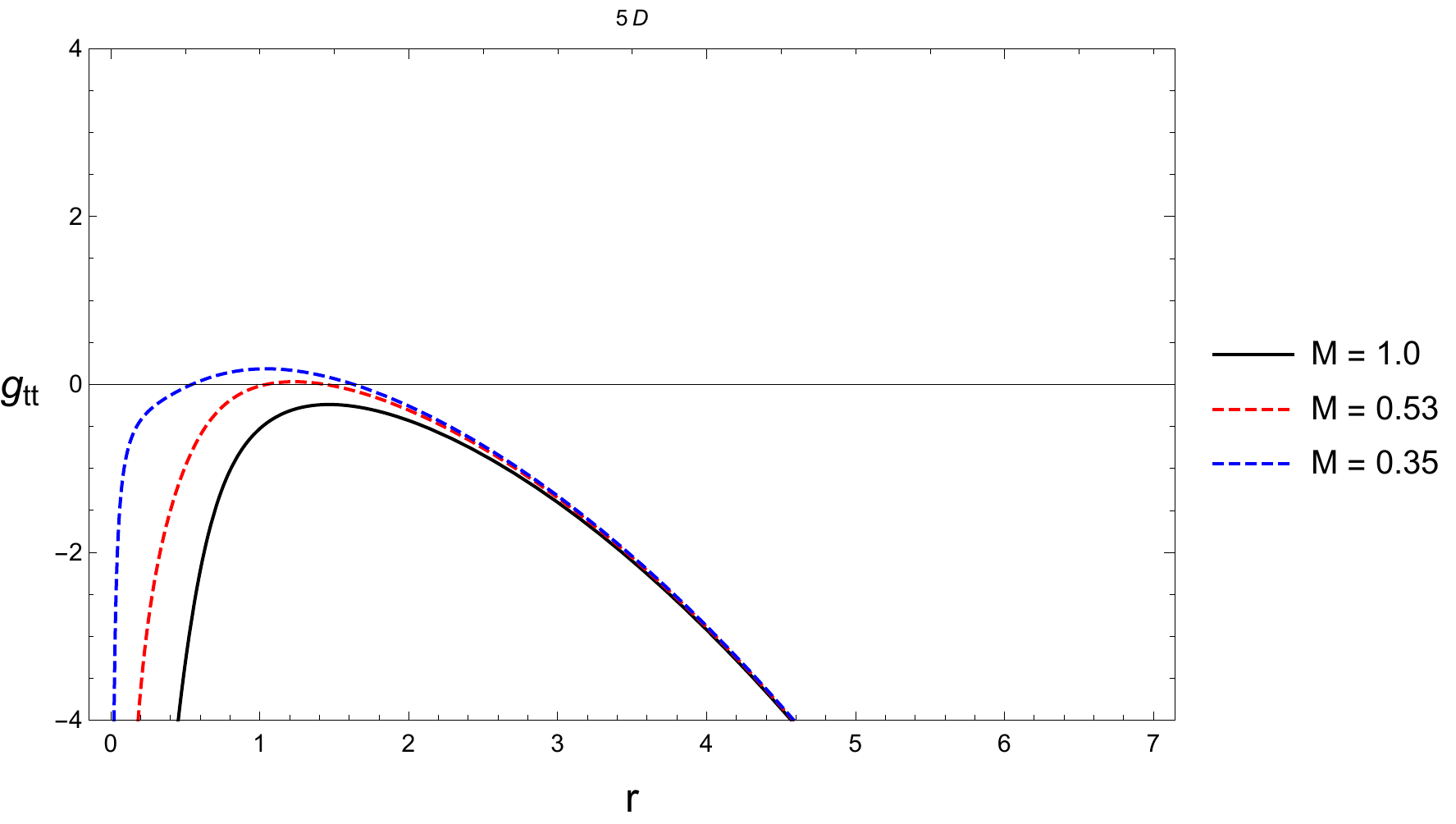}
	\caption {A typical plot of $5d$ $ g_{tt} $ with $ \kappa=0.3$, $ q=0.5 $ and $ \Lambda=2$. There are at most two horizons. The exact numerical values of the radii are shown in Table~\ref{table:4}.}
	\label{fig:gtt5ds}
\end{figure}
\begin{equation}
B(r)_{5}= 1+ \frac{4 \kappa q^2}{\lambda r^{6}}.
\end{equation}
\begin{table}[h!]
	\centering
	\begin{tabular}{||c c c c ||} 
		\hline
		$ M $ & $ r_{h1} $& $ r_{e} $  & $ r_{h2} $  \\ [0.5ex] 
		\hline\hline
		1 & - & - & - \\ 
		0.53 & - & 1.05 & - \\
		0.35 & 0.55 & - & 1.62 \\ [1ex] 
		\hline
	\end{tabular}
	\caption{An exact horizon $ r_{h} $ and extremal $ r_{e} $ radius from Fig.~\ref{fig:gtt5ds}.}
	\label{table:4}
\end{table}

\begin{figure}[htbp]
	\centering\leavevmode
	\epsfysize=7.5cm \epsfbox{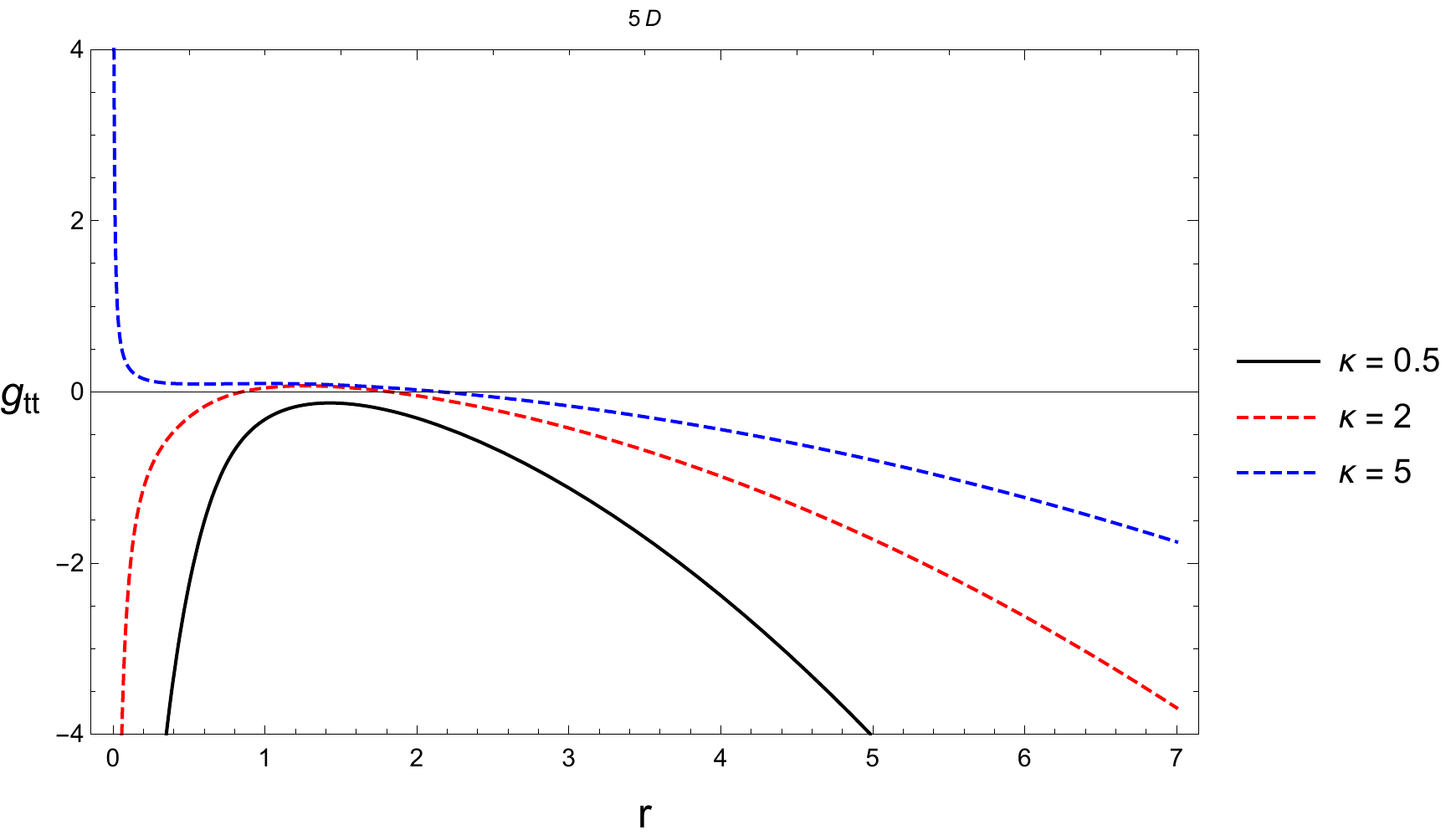}
	\caption {A typical plot of $5d$ $ g_{tt} $ with $ M=1$, $ q=0.5 $ and $ \Lambda=2$.  The exact numerical values of the radii are shown in Table~\ref{table:5}.}
	\label{fig:gtt5dskappa}
\end{figure}
\begin{table}[h!]
	\centering
	\begin{tabular}{||c c c ||} 
		\hline
		$ \kappa $ & $ r_{h1} $& $ r_{h2} $ \\ [0.5ex] 
		\hline\hline
		0.5 & - & - \\ 
		2 & 0.85 & 1.81 \\
		5 & - & 2.15 \\ [1ex] 
		\hline
	\end{tabular}
	\caption{An exact horizon radius $ r_{h} $ from Fig.~\ref{fig:gtt5dskappa}.}
	\label{table:5}
\end{table}

Typical metric solutions $g_{tt}$ are shown in Figs.~\ref{fig:gtt5ds}-{\ref{fig:gtt5dskappa}. For dS solutions there are at most two horizons, just like the case of Maxwell electrodynamics. In Fig.~\ref{fig:gtt5ds} we vary the black hole mass to show the existence of solutions with two, one, and no (or naked singularity) horizon. Fig.~\ref{fig:gtt5dskappa} shows variations of solutions for different $\kappa$ with fixed mass. These profiles are qualitatively generic for $D\geq5$.

\subsubsection{$D=6$}

In $6d$, we have 
\begin{equation}
\label{131}
V(\bar{r})= \sqrt{\frac{1}{\lambda }-\frac{\kappa  q^2}{\bar{r}^8}},
\end{equation}
\begin{equation}
\label{132}
U(\bar{r}) =  \left(\frac{1}{\lambda}+\frac{\kappa q^2}{\bar{r}^{8}} \right) V(\bar{r})^{-1},
\end{equation}
and
\begin{eqnarray}
\label{133}
e^{2\psi(\bar{r})}= 1-\frac{2M}{\bar{r}^{3}}-\frac{\bar{r}^{2}}{5\kappa}  + \frac{\bar{r}^2 \,
	_2F_1\left(-\frac{1}{2},-\frac{5}{8};\frac{3}{8};q^2 \bar{r}^{-8} \kappa  \lambda \right)}{5\lambda^{\frac{1}{2}}\kappa}. 
\end{eqnarray}
Transforming it, as before, into Tangherlini gauge $r^2 =\left(\sqrt{\frac{1}{\lambda }-\frac{\kappa  q^2}{\bar{r}^8}}\right)\bar{r}^2$ we get
\begin{equation}
\label{134}
ds^2 = - A(r) f(r) dt^2 + \frac{ r^{2} \sqrt{\frac{\lambda}{2}} A(r) \sqrt{1+\sqrt{B(r)_6}}}{ B(r)_6 f(r)} dr^2 + r^2 d\Omega^2,
\end{equation}
with
\begin{equation}
A(r)=  \frac{ \frac{1}{\lambda} + \frac{4\kappa q^2}{\lambda^2 r^8 \left(1+\sqrt{B(r)_6}\right)^2}  }{  \sqrt{\frac{1}{\lambda} - \frac{4\kappa q^2}{\lambda^2 r^8 \left(1+\sqrt{B(r)_6}\right)^2}}},
\end{equation}
\begin{eqnarray}
\label{135}
f(r)=  1-\frac{2M}{y(r)^{3}}-\frac{y(r)^{2}}{5\kappa}  + \frac{y(r)^2 \,
	_2F_1\left(-\frac{1}{2},-\frac{5}{8};\frac{3}{8};q^2 y(r)^{-8} \kappa  \lambda \right)}{\lambda^{\frac{1}{2}} 5 \kappa},
\end{eqnarray}
\begin{equation}
y(r)=\left({\lambda \over 2}\right)^{\frac{1}{4}} \left(1+\sqrt{B(r)_{6}}\right)^{\frac{1}{4}}\ r,
\end{equation}
and
\begin{equation}
B(r)_{6}= 1+ \frac{4 \kappa q^2}{\lambda r^{8}}.
\end{equation}




\begin{figure}[htbp]
	\centering\leavevmode
	\epsfysize=7.5cm \epsfbox{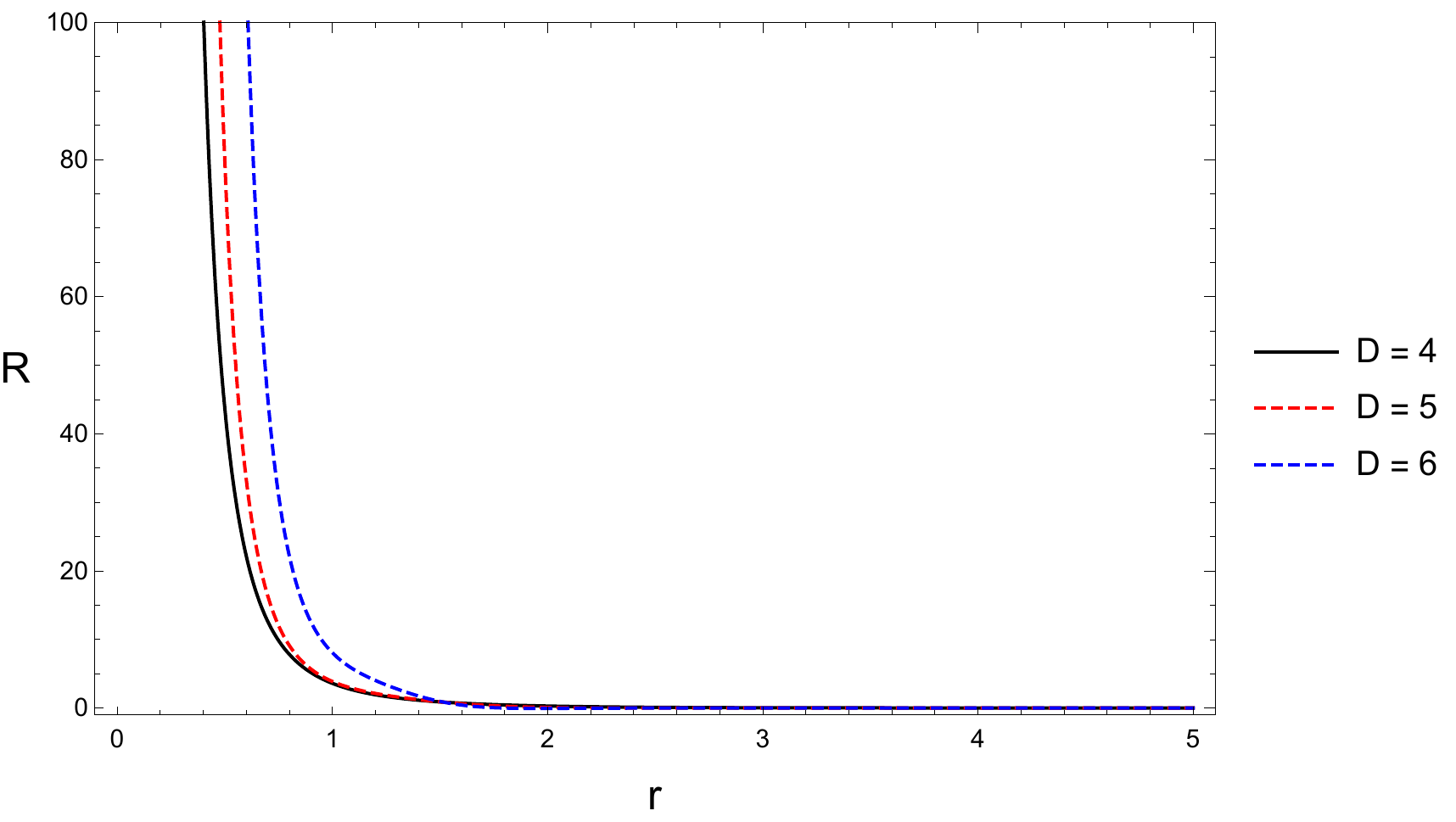}
	\caption {The Ricci scalar $ \mathcal{R} $ with $ M=1 $, $ q=1 $ and $ \kappa=10.0 $.}
	\label{fig:ricci}
\end{figure}
In all dimensions, the ``physical" Ricci scalars suffer from singularity at the origin, as shown in Fig.~\ref{fig:ricci}. Our solutions, thus, can be interpreted as genuine black holes. 


\subsection{$\tilde{\alpha}=2$}

Another condition we can investigate is when $\tilde{\alpha}=2$. In this constraint, Eq.~\eqref{115} reduces to 

\begin{equation}
V(\bar{r})= \left(\frac{1}{\lambda ^2}-\frac{2 \kappa  q^2 }{\lambda \bar{r}^{2 (D-2)} }\right)^{\frac{1}{D-2}}
\end{equation}
\begin{equation}
U(\bar{r})= \frac{\sqrt{  \frac{1}{\lambda^2} - \frac{2\kappa q^2}{\lambda \bar{r}^{2(D-2)}}  }}{\lambda \left(   \frac{1}{\lambda^2} - \frac{2\kappa q^2}{\lambda \bar{r}^{2(D-2)}}  \right)},
\end{equation}
\begin{eqnarray}
e^{2\psi(\bar{r})}&=& 1- \frac{2M}{\bar{r}^{D-3}} -\frac{\bar{r}^2}{(D-1) \kappa } + \frac{\bar{r}^2 \left(1-2 \kappa  \lambda  q^2 \bar{r}^{4-2 D}\right)^{\frac{1}{2-D}} \left(\frac{1-2 \kappa  \lambda  q^2 \bar{r}^{4-2 D}}{\lambda ^2}\right)^{\frac{1}{D-2}}}{(D-1) \kappa } \nonumber \\ && \times \,
_2F_1\left(\frac{1}{2-D},-\frac{D-1}{2 (D-2)};\frac{D-3}{2 (D-2)};2 q^2 \bar{r}^{4-2 D} \kappa  \lambda \right).
\end{eqnarray}
	
Under the Tangherlini transformation $ r^2 = \left(\frac{1}{\lambda ^2}-\frac{2 \kappa  q^2 }{\lambda \bar{r}^{2 (D-2)} }\right)^{\frac{1}{D-2}} \bar{r}^2 $ they yield
\begin{equation}
\bar{r}(r)= \left( \lambda^2 r^{2(D-2)} + 2\kappa q^2 \lambda  \right)^{\frac{1}{D-2}},
\end{equation}

	\begin{equation}
	V(r) = \left(\frac{\lambda -\frac{2 \kappa  q^2}{\left(\lambda  r^{2 D-4}+2 \kappa  q^2\right)^2}}{\lambda ^3}\right)^{\frac{1}{D-2}},
	\end{equation}

\begin{equation}
U(r) =\frac{\sqrt{  \frac{1}{\lambda^2} - \frac{2\kappa q^2}{\lambda   	\left( \lambda^2 r^{2(D-2)} + 2\kappa q^2 \lambda  \right)^{(D-2)}}  }}{\lambda \left(   \frac{1}{\lambda^2} - \frac{2\kappa q^2}{\lambda 	\left( \lambda^2 r^{2(D-2)} + 2\kappa q^2 \lambda  \right)^{(D-2)}}  \right)},
	\end{equation}
and	
	\begin{eqnarray}
	e^{2\psi(r)}& =& 1-  2 M \left(\lambda  \left(\lambda  r^{2 D-4}+2
	\kappa  q^2\right)\right)^{\frac{1}{D-2}-1}-\frac{\left(\lambda  \left(\lambda  r^{2 D-4}+2 \kappa  q^2\right)\right)^{\frac{2}{D-2}}}{(D-1) \kappa } \nonumber \\ && + \frac{1}{(D-1) \kappa } \bigg[  \left(\frac{\lambda -\frac{2 \kappa  q^2}{\left(\lambda  r^{2 D-4}+2 \kappa  q^2\right)^2}}{\lambda ^3}\right)^{\frac{1}{D-2}} \left(1-\frac{2 \kappa  q^2}{\lambda 
			\left(\lambda  r^{2 D-4}+2 \kappa  q^2\right)^2}\right) \nonumber \\ && \times \left(\lambda  \left(\lambda  r^{2 D-4}+2 \kappa  q^2\right)\right)^{\frac{2}{D-2}} \, _2F_1\left(1,\frac{D-1}{2
			(D-2)};\frac{D-3}{2 (D-2)};\frac{2 q^2 \kappa }{\lambda  \left(\lambda  r^{2 D-4}+2 q^2 \kappa \right)^2}\right) \bigg]. \nonumber \\ 
	\end{eqnarray}

\section{Thermodynamics of EiBI Black Holes}
\label{sec:thermo}

As is well known, black holes are thermodynamical objects that radiate and can undergo phase transitions~\cite{Bardeen:1973gs, Bekenstein:1973ur}. While there has been an extensive study on their thermodynamical properties in GR's framework (for example, see~\cite{Roychowdhury:2014cva} and references therein), there has been no comprehensive investigation in EiBI gravity. To the best of our knowledge, such attempts to study the entropy of modified, in particular the Born-Infeld types of, gravity were done by the authors in~\cite{He:2016yuc, Ozen:2017uoe}. In this work we try to investigate the thermodynamical stability of the solutions we obtain in the previous sections.

We start from the Hawking temperature~\cite{Hawking:1974sw},
\begin{equation}
\label{tau}
T_H= \frac{\tau}{2 \pi},
\end{equation}
where $ \tau $ is the surface gravity, given by~\cite{Visser:1992qh}:
\begin{equation}
\label{temperaturumum}
\tau\equiv\frac{\partial_{r} g_{tt}}{2\sqrt{g_{tt}g_{rr}}}\bigg|_{r=r_+},
\end{equation}
with $r_+$ the black hole's (outer) horizon. Note that this is a general form of surface gravity that is more suitable when $g_{rr}\neq g_{tt}^{-1}$. 
Then, we can calculate the specific heat
\begin{equation}
\label{c4}
C_q =  T_H \left(    \frac{\partial S}{\partial T_H}\right)_q
\end{equation}
where it is sought under the constant charge. In this section, we would present these thermodynamical quantities in Maxwell and Born-Infeld electrodynamics and investigate their stability. It was shown by Padmanabhan~\cite{Kothawala:2007em,Padmanabhan:2002sha,Paranjape:2006ca} that  the Einstein's equations near the horizon satisfy the first law of black hole thermodynamics. In~\cite{He:2016yuc} the authors show that such first law is still obeyed by EiBI gravity. Here we shall show that higher-dimensional EiBI black holes still obey the first law, and give the higher-dimensional corrected entropy.

\subsection{EiBI-Maxwell}

For the EiBI metric~\eqref{5} the Hawking temperature reads ($f(r_+)=0$)
\begin{equation}
T_H(r_+)=\frac{\psi(r_+) f'(r_+)}{4\pi}.
\end{equation}
From solutions~\eqref{36a} and \eqref{42}	 it takes
\begin{eqnarray}
\label{maxwelltemp}
T_H &=&  \frac{\left(r_+^{2 D-4}+\frac{\kappa  q^2}{\lambda }\right)^{\frac{1}{2-D}}}{8 \pi  (D-2) (D-1) \kappa  \lambda ^2 r_+ \left(\kappa  q^2 r_+^4-\lambda  r_+^{2 D}\right)^2} \bigg[ 2 (D-3) (D-2) \kappa ^2 \lambda  q^2 r_+^{6-2 D} \nonumber \\ && \times \left((D+1) \kappa ^2 q^4 r_+^8+2 (D-3) \kappa  \lambda  q^2 r_+^{2 D+4}+(D-3) \lambda ^2 r_+^{4 D}\right) \mathcal{K}_1\nonumber \\ &&  + (D+1) r_+^{-4D}  \bigg[(D-3) \kappa  r_+^2 ~\Gamma \left(\frac{D-1}{2 (D-2)}\right)  \bigg[ -2 \kappa  \lambda  q^2 r_+^{2 D+4} \nonumber \\ && \times \left( (4-D) \kappa ^2 q^4 r_+^8+2 (D-3) \kappa  \lambda  q^2 r_+^{2 D+4}+(D-2) \lambda ^2 r_+^{4 D}\right)  \mathcal{K}_2\nonumber \\ && + 2 (D-2) \lambda ^2 r_+^{4 D} \left(\kappa  q^2 r_+^4-\lambda  r_+^{2 D}\right)^2 \mathcal{K}_3 - \kappa  q^2 r_+^4 \left(\kappa ^2 q^4 r_+^8-\lambda ^2 r_+^{4 D}\right) \bigg[(3 D-7) \kappa  q^2 r_+^4 \mathcal{K}_4 \nonumber \\ && + 2 (D-2) \lambda  r_+^{2 D} \mathcal{K}_5    \bigg] \bigg]+  \left(\kappa  q^2 r_+^4-\lambda  r_+^{2 D}\right) \bigg[ 2(D-2) \bigg[ \kappa  q^2 r_+^8 ~\Gamma \left(\frac{7-3 D}{4-2 D}\right) \nonumber \\ &&  +\lambda ^2 r_+^{4 D} \left(\lambda  r_+^{2 D}-\kappa  q^2 r_+^4\right) \left(\lambda  r_+^{2 D-4}+\kappa 
	q^2\right)^{\frac{2}{D-2}} \bigg[ \lambda  r_+^{2 D} \left(\lambda  r_+^{2 D}+3 \kappa  q^2 r_+^4\right) \mathcal{K}_6   \nonumber \\ &&  -    3 \kappa  q^2 r_+^4 \left(\lambda  r_+^{2 D}+\kappa  q^2 r_+^4\right) \mathcal{K}_7  \bigg] \bigg] +  \lambda  r_+^{2 D+4} ~\Gamma \left(-\frac{D-1}{2 (D-2)}\right) \nonumber \\ && \times \bigg[ \lambda  r_+^{2 D} \left((5-D) \kappa  q^2 r_+^4+(D-1) \lambda  r_+^{2 D}\right)  \mathcal{K}_8   +   (D-5) \kappa  q^2 r_+^4 \left(\lambda  r_+^{2 D}+\kappa  q^2 r_+^4\right) \mathcal{K}_9 \bigg] \bigg] \bigg] \bigg],\nonumber \\ 
\end{eqnarray}
	where
	\begin{equation}
	\mathcal{K}_1\equiv\, _2F_1\left(1,\frac{7-3 D}{4-2 D};\frac{1}{2} \left(\frac{1}{D-2}+3\right);-\frac{\kappa  q^2 r_+^{4-2 D}}{\lambda }\right),   
	\end{equation}
	\begin{equation}
	\mathcal{K}_2\equiv\, _2\tilde{F}_1\left(1,\frac{7-3 D}{4-2 D};\frac{1}{2} \left(\frac{1}{D-2}+3\right);-\frac{\kappa  q^2 r_+^{4-2 D}}{\lambda }\right)  ,
	\end{equation}
	\begin{equation}
	\mathcal{K}_3\equiv\, _2\tilde{F}_1\left(1,\frac{D-3}{2 (D-2)};\frac{D-1}{2 (D-2)};-\frac{\kappa  q^2 r_+^{4-2 D}}{\lambda }\right)  ,
	\end{equation}
	\begin{equation}
	\mathcal{K}_4\equiv\, _2\tilde{F}_1\left(2,\frac{1}{4-2 D}+\frac{5}{2};\frac{1}{2} \left(\frac{1}{D-2}+5\right);-\frac{\kappa  q^2 r_+^{4-2 D}}{\lambda }\right),
	\end{equation}
	\begin{equation}
	\mathcal{K}_5\equiv\, _2\tilde{F}_1\left(2,\frac{7-3 D}{4-2 D};\frac{1}{2} \left(\frac{1}{D-2}+3\right);-\frac{\kappa  q^2 r_+^{4-2 D}}{\lambda }\right),
	\end{equation}
	\begin{equation}
	\mathcal{K}_6\equiv\, _2\tilde{F}_1\left(1,\frac{3 (D-3)}{2 (D-2)};\frac{7-3 D}{4-2 D};-\frac{\kappa  q^2 r_+^{4-2 D}}{\lambda }\right),
	\end{equation}
	\begin{equation}
	\mathcal{K}_7 = \, _2\tilde{F}_1\left(2,\frac{3}{4-2 D}+\frac{5}{2};\frac{1}{4-2 D}+\frac{5}{2};-\frac{\kappa  q^2 r_+^{4-2 D}}{\lambda }\right),
	\end{equation}
	\begin{equation}
	\mathcal{K}_8 = \, _2\tilde{F}_1\left(1,\frac{D-5}{2 (D-2)};\frac{D-3}{2 (D-2)};-\frac{\kappa  q^2 r_+^{4-2 D}}{\lambda }\right),
	\end{equation}
	\begin{equation}
	\mathcal{K}_9 = \, _2\tilde{F}_1\left(2,\frac{3 (D-3)}{2 (D-2)};\frac{7-3 D}{4-2 D};-\frac{\kappa  q^2 r_+^{4-2 D}}{\lambda }\right).
	\end{equation}

As in the case of the its corresponding metric solutions, the temperature above looks unappealing. For a better look, we present the result in various dimensions.


In $D=4$ the temperature~\eqref{maxwelltemp} is given by 
\begin{equation}
\label{t4}
T_H=\frac{\left(r^2_+-q^2-\Lambda  r^4_+ \right) \sqrt{\frac{\lambda  r^4_+}{\kappa  q^2+\lambda  r^4_+}}}{4 \pi  r^3_+}.
\end{equation}
One can see that as $\kappa\rightarrow0$ the temperature reduces to that of Reissner-Nordstrom black hole (for example, see~\cite{Chamblin:1999tk, Chamblin:1999hg}).

To study the nature of the temperature in detail, we plot the $ T_H $ vs $ r_+ $ in Fig.~\ref{fig:t4fig}. 
\begin{figure}[htbp]
	\centering\leavevmode
	\epsfysize=6cm \epsfbox{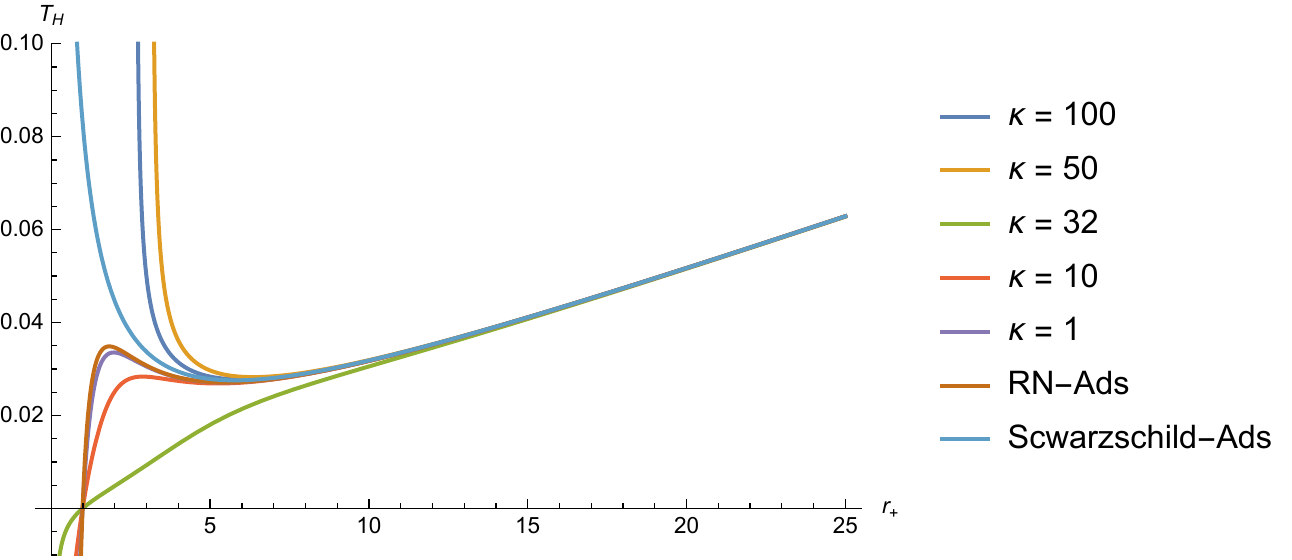}
	\caption {A typical plot of $4d$ $ T_{H} $ in equation \eqref{t4}  with $ q=1 $ and $ \Lambda=-0.03 $.}
	\label{fig:t4fig}
\end{figure}
An interesting thing here is that there is a discontinuous transition of the temperature plot from lower to higher $\kappa$. As is known, and can be seen from the plot, there are two local optima in the RN-AdS temperature; the stable and the unstable ones. as the $\kappa$ value is turned on we see the transition that brings the two to merge. At $\kappa=32$ (with $q=1$ and $\Lambda=-0.03$) the local optima disappear and the temperature plot seems to be almost linear. However, when $\kappa$ is tuned to higher value, for example $\kappa=50$, then it abruptly switches the qualitative shape of the plot and approaches the Schwarzschild-AdS temperature. This discontinuity is the sign of $\kappa$ screening effect; {\it i.e.}, the strength of $\kappa$ effectively ``screens" the charge. 

	
Interestingly, such behavior seems to be unique feature of $4d$ black hole. In $D>4$ the temperature behaves quite distinctly. In $D=5$, for example, the temperature reads. 
\begin{equation}
\label{t5}
T_H = \frac{-\kappa  q^2 \lambda^{\frac{1}{3}}+r_+^4 \lambda^{\frac{1}{3}}\sqrt[3]{\kappa  q^2+\lambda  r_+^6}+2 \kappa \lambda^{\frac{1}{3}}  r_+^2 \sqrt[3]{\kappa  q^2+\lambda  r_+^6}-\lambda^{\frac{4}{3}}  r_+^6}{4 \pi  \kappa  r_+ \left(\kappa  q^2\lambda
	+\lambda r_+^6 \right)^{\frac{2}{3}} }.
\end{equation}
\begin{figure}[htbp]
	\centering\leavevmode
	\epsfysize=6cm \epsfbox{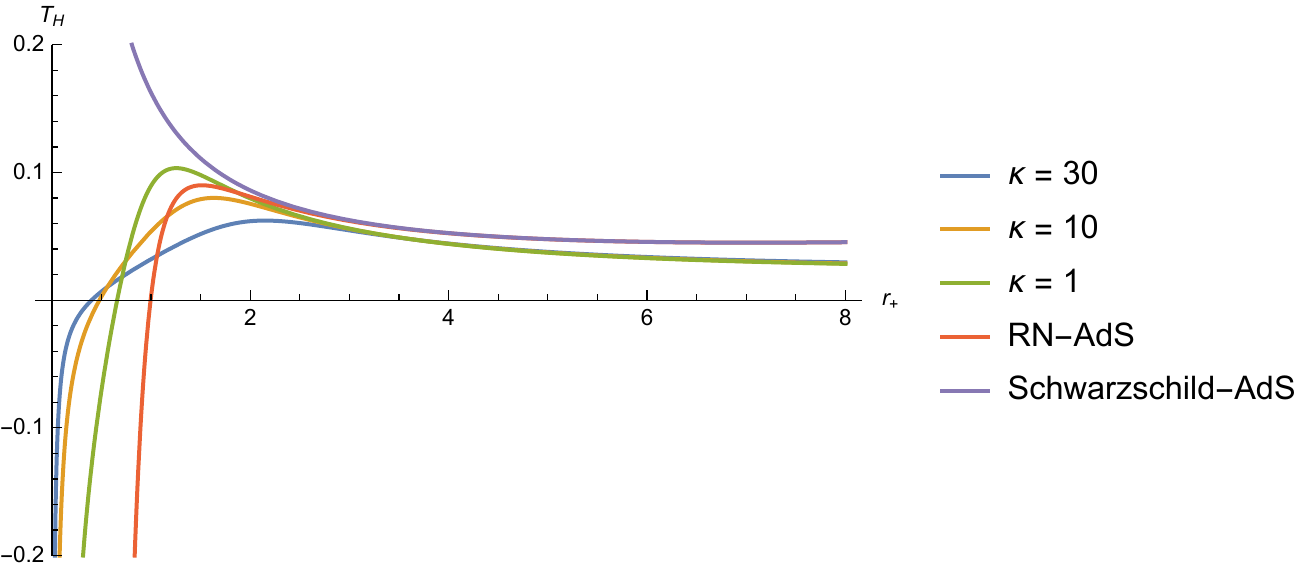}
	\caption {A typical plot of $5d$ $ T_{H} $ in equation \eqref{t5}  with $ q=1 $ and $ \Lambda=-0.03 $.}
	\label{fig:t5fig}
\end{figure}
In Fig.~\ref{fig:t5fig}, the story is the same as in Fig.~\ref{fig:t4fig} up to $\kappa=30$. Beyond that, no (real) temperature exists. There is no discontinuous transition from RN-AdS-like to Schwarzschild-AdS. 

	


To ensure these solutions satisfy the first law of thermodynamics, we should start with Eq.~\eqref{35} at $r=r_+$,
\begin{equation}
\label{1law}
r_{+} f' - (D-3)= \frac{r_{+}^2}{\kappa} \left[ 1- \left(\lambda +\frac{\kappa q^2}{r_{+}^{2(D-2)}}\right)^{2 \over (D-2)}\right].
\end{equation}
The second term on the right-hand side can be expanded binomially in powers of $ \kappa $,
\begin{eqnarray}
\left(\lambda +\frac{\kappa q^2}{r_{+}^{2(D-2)}}\right)^{2 \over (D-2)}&=&\lambda^{\frac{2}{D-2}} + \sum_{i=1}^{\infty} (-1)^{i+1}~ \frac{a_{i}~ \kappa^{i} q^{2i}~ \lambda^{-i+\frac{2}{D-2}}}{r^{2i(D-2)}_{+}},\nonumber\\
&=&\lambda^{\frac{2}{D-2}}+\frac{2 \kappa  q^2 \lambda ^{\frac{2}{D-2}-1} r^{4-2 D}}{D-2}-\frac{(D-4) \kappa ^2 q^4 \lambda ^{\frac{2}{D-2}-2} r^{8-4 D}}{(D-2)^2} \nonumber\\&&+ \frac{2 (D-4) (D-3) \kappa ^3 q^6 \lambda ^{\frac{2}{D-2}-3} r^{12-6 D}}{3 (D-2)^3}+\mathcal{O}[\kappa]^{4},
\end{eqnarray} 
with $a_i$ are the series' coefficients.

Eq.~\eqref{1law} can then be written as
\begin{equation}
\label{thermo1law}
T~ \frac{2 \pi r^{D-3}_{+}}{\psi} dr_{+} - d\left( \frac{r^{D-3}_+}{2} + \sum_{i=1}^{\infty} (-1)^{i+1}~ \frac{\alpha_{i}~ \kappa^{i-1} q^{2i}~ \lambda^{-i+\frac{2}{D-2}}}{r^{2i(D-2)+1-D}_{+}} \right) = P_D dV,
\end{equation}
where the first three values of the coefficient $\alpha_i$'s are shown below
\begin{align}
\label{alpha}
\alpha_{1}= \frac{1}{(D-2)(D-3)},~~ \alpha_2 = \frac{(D-4)}{2(D-2)^{2}~(3D-7)},~~ \alpha_3 = \frac{(D-4)(D-3)}{3(D-2)^3 ~(5D-11)},
\end{align}
and $P_D\equiv -\frac{1}{8\pi} \left(\frac{\lambda^{\frac{2}{D-2}}-1}{\kappa}\right)$ and $V\equiv \frac{\omega_{D-2} r_+^{D-1}}{D-1}$. We can thus infer the black hole mass $M(r_+)$,
\begin{equation}
M(r_+)\equiv\frac{r^{D-3}_+}{2} + \sum_{i=1}^{\infty} (-1)^{i+1}~ \frac{\alpha_{i}~ \kappa^{i-1} q^{2i}~ \lambda^{-i+\frac{2}{D-2}}}{r^{2i(D-2)+1-D}_{+}}.
\end{equation}
In the lowest order, the mass reduces to the RN-like in $D$ dimensions,
\begin{equation}
M(r_+)= \frac{r_{+}^{D-3}}{2} +\frac{q^2 ~\lambda^{\frac{4-D}{D-2}}}{(D-2)(D-3)~r^{D-3}}.
\end{equation} 
Eq.~\eqref{thermo1law} shows that at the horizon the EiBI black holes satisfy the first law of black hole thermodynamics. The entropy can be extracted as
\begin{eqnarray}
S&=&\int \frac{2\pi r_{+}^{D-3}}{\psi(r_{+})} dr_{+},\nonumber\\
 &=& \frac{2\pi r_{+}^{D-2}}{(D-2)}\,_{2}F_{1}\left(-\frac{1}{2},-\frac{1}{(D-2)},\frac{1}{2},-\frac{\kappa q^2 r_{+}^{4-2D}}{\lambda}\right).
\end{eqnarray} 
It reduces back to the $ 4d $ GR entropy as $ \kappa \rightarrow 0 $, $ S_{BH}= \pi r_{+}^2 $.

The heat capacity is calculated below. In $4d$ it reads
\begin{equation}
\label{ceibi4}
C_{q}= \frac{2 \pi  \left(q^2+\Lambda  r_+^4-r_+^2\right) \left(\kappa ^2 q^4-\lambda ^2 r_+^8\right)}{\sqrt{\lambda}r_+^2 \left(\kappa  q^4+q^2 \left(3 r_+^4 (\lambda -\kappa  \Lambda )+\kappa  r_+^2\right)-\lambda  r_+^6 \left(\Lambda  r_+^2+1\right)\right)}
\end{equation}
The graph of the heat capacity shown in Fig.~\ref{fig:ceibi4fig} is varied by EiBI parameter $ \kappa $. It implies that we have two phase of negative and positve heat capacities. For EiBI framework when $ \kappa=1 $ and $ \kappa=10 $, it is shown that the black hole is stable at smaller radius than RN-AdS.
\begin{figure}[htbp]
	\centering\leavevmode
	\epsfysize=6cm \epsfbox{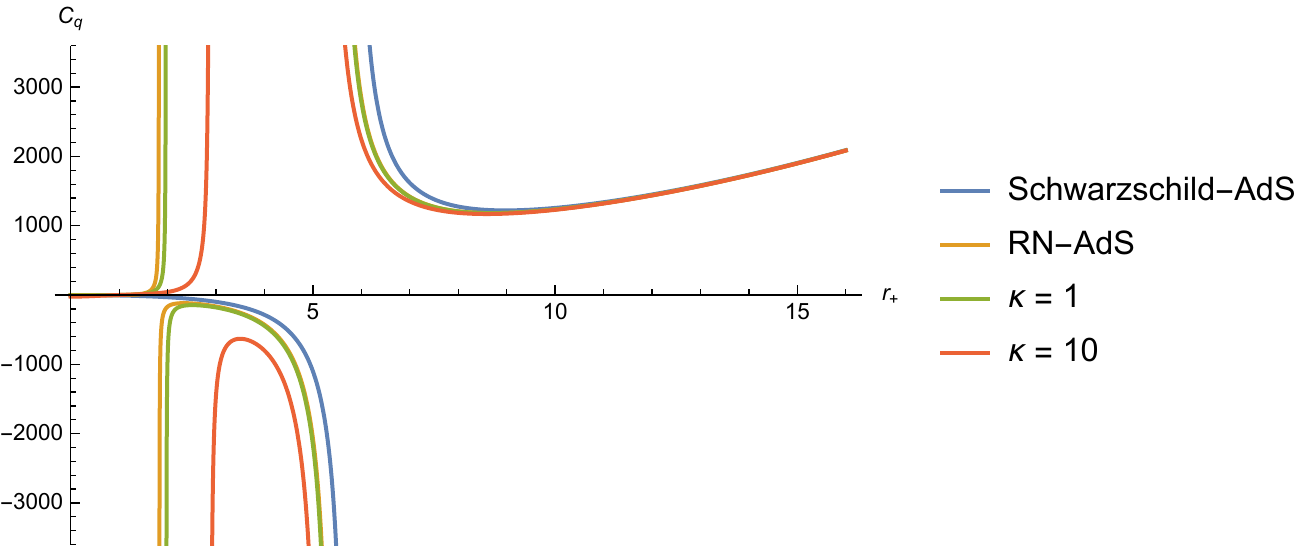}
	\caption {A typical plot of $4d$ $ C_q $ in equation \eqref{ceibi4}  with $ q=1 $ and $ \Lambda=-0.03 $.}
	\label{fig:ceibi4fig}
\end{figure} 
In $ D = 5 $, 
\begin{equation}
\label{ceibi5}
C_q =-\frac{2 \pi  r_+^3 \sqrt[3]{\frac{\kappa  q^2}{\lambda  r_+^6}+1} \left(\kappa  q^2+\lambda  r_+^6\right) \left(\kappa  q^2-r_+^4 \sqrt[3]{\kappa  q^2+\lambda  r_+^6}-2 \kappa  r_+^2
	\sqrt[3]{\kappa  q^2+\lambda  r_+^6}+\lambda  r_+^6\right)}{\kappa ^2 q^4+\kappa  q^2 r_+^2 \left(2 \kappa +3 r_+^2\right) \sqrt[3]{\kappa  q^2+\lambda  r_+^6}+\lambda  r_+^8 \left(r_+^2-2
	\kappa \right) \sqrt[3]{\kappa  q^2+\lambda  r_+^6}-\lambda ^2 r_+^{12}}
\end{equation}
Corresponding to $ 5d $ temperature, the discontinuity makes no transition from RN-AdS-like to Schwarzschild-AdS-like. Hence, from the heat capacity point of view in Fig.~\ref{fig:ceibi5fig}, there is no transition from small black hole to the large black hole.

\begin{figure}[htbp]
	\centering\leavevmode
	\epsfysize=6cm \epsfbox{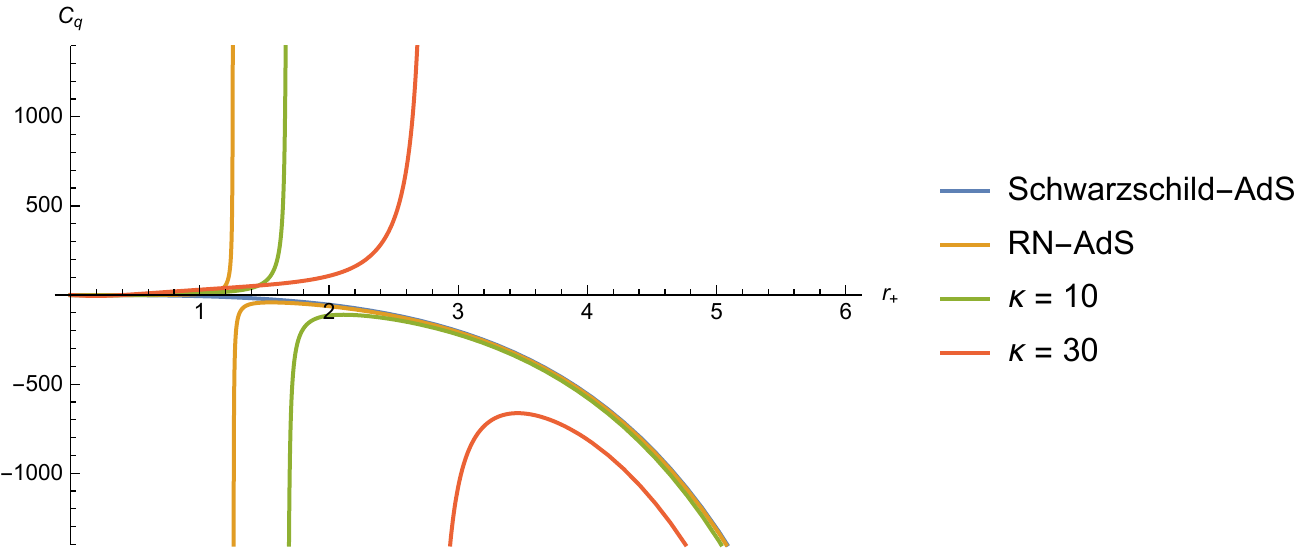}
	\caption {A typical plot of $5d$ $ C_q $ in equation \eqref{ceibi5}  with $ q=1 $ and $ \Lambda=-0.03 $.}
	\label{fig:ceibi5fig}
\end{figure}

\subsection{Born-Infeld}

For the metric~\eqref{56} the Hawking temperature reads
\begin{equation}
T(\bar{r}_+)= \frac{1}{4\pi} \left( \frac{U'}{U} e^{2\psi} + (e^{2\psi})'\right) = \frac{(e^{2\psi})'}{4\pi}.
\end{equation}

For $\alpha=1$ the corresponding temperature is, 

\begin{eqnarray}
T_H &=& \frac{1}{\pi  (\rho(r_+)-1) (D-1)^2 \kappa  r_+ \left(\kappa  q^2 \eta(r_+)^{4-2 D}+\frac{1}{\lambda }\right) \left(\eta(r_+)^{2 D}-16^{\frac{1}{2-D}} \kappa  q^2 r_+^4 \rho(r_+)^{\frac{4}{D-2}} \lambda^{\frac{D+2}{D-2}}\right)^2} \nonumber \\ && \times \bigg[ 2^{-\frac{2 (D+4)}{D-2}} \left(\frac{1}{\lambda }-\kappa  q^2 \eta(r_+)^{4-2 D}\right)^{\frac{D-4}{D-2}} \left(\frac{1}{\lambda }-16^{\frac{1}{2-D}} \kappa  q^2 r_+^4
(\rho(r_+)+1)^{\frac{4}{D-2}} \eta(r_+)^{-2 D} \lambda ^{\frac{4}{D-2}}\right)^{\frac{2}{D-2}} \nonumber \\ && \times \bigg[ 2^{\frac{10}{D-2}} (D-1)^2 \eta(r_+)^{4 D} \bigg[ 2^{\frac{2}{D-2}} (D-3) \kappa -r_+^2 \rho(r_+)^{\frac{2}{D-2}} \bigg[\lambda ^{\frac{2}{D-2}} \nonumber \\ && -\left(1-16^{\frac{1}{2-D}} \kappa  q^2 r_+^4 \rho(r_+)^{\frac{4}{D-2}} \eta(r_+)^{-2 D} \lambda
^{\frac{D+2}{D-2}}\right)^{\frac{2}{D-2}}\bigg]+(D-1) \kappa ^2 q^4 r_+^8 \rho(r_+)^{\frac{8}{D-2}} \lambda ^{\frac{6}{D-2}} \nonumber \\ && \times \bigg[ r_+^2 \rho(r_+)^{\frac{2}{D-2}}-2^{\frac{4}{D-2}} (D-3) (D-1) \kappa  \lambda ^{\frac{2}{D-2}+2} \bigg[ \left(4^{\frac{1}{D-2}} (D+1)-2^{\frac{D}{D-2}}\right) \lambda ^{\frac{2 D}{D-2}} \nonumber \\ && -2^{\frac{2}{D-2}} (D-1) \lambda ^{\frac{2}{D-2}+2} \left(1-16^{\frac{1}{2-D}} \kappa  q^2 r_+^4 \rho(r)^{\frac{4}{D-2}} \eta(r_+)^{-2 D} \lambda ^{\frac{D+2}{D-2}}\right)^{\frac{2}{D-2}} \bigg]   \bigg]     \bigg] \bigg]\bigg],
\end{eqnarray}
where 
\begin{equation}
\eta(r_+)\equiv 2^{\frac{1}{2-D}} r_+ \left( \sqrt{\frac{4 \kappa  q^2 r_+^{4-2 D}}{\lambda }+1}+1  \right)^{\frac{1}{D-2}} \lambda ^{\frac{1}{D-2}},
\end{equation}
and 
\begin{equation}
\rho(r_+)\equiv \sqrt{\frac{4 \kappa  q^2 r_+^{4-2 D}}{\lambda }+1}+1.
\end{equation}

In particular, the $4d$ temperature reads
\begin{equation}
\label{t4bi}
T_H = \frac{\lambda  r_+ \left(-2 \kappa  q^2-r_+^2 \left((\lambda -1) r_+^2-\kappa \right) \left(\sqrt{\frac{4 \kappa  q^2}{\lambda  r_+^4}+1}+1\right)\right)}{4 \pi  \kappa  \left(4 \kappa 
	q^2+\lambda  r_+^4 \left(\sqrt{\frac{4 \kappa  q^2}{\lambda  r_+^4}+1}+1\right)\right)}.
\end{equation}
\begin{figure}[htbp]
	\centering\leavevmode
	\epsfysize=6cm \epsfbox{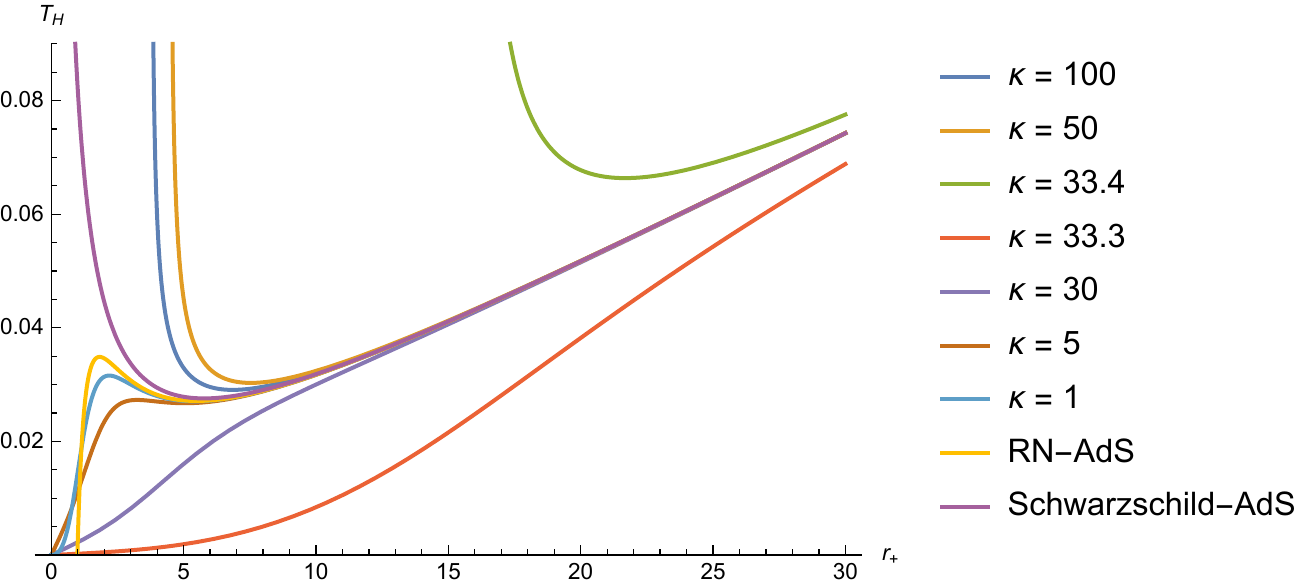}
	\caption {A plot of $4d$ EiBI-BI $T_{H} $ with $ q=1 $ and $ \Lambda=-0.03 $.}
\label{fig:t4bornfig}
\end{figure}
In Fig~\ref{fig:t4bornfig}, we show a profile of $4d$ $ T_H $ as a function or $ r_+ $ for several $ \kappa $ values. there are several interesting features here. First, unlike its Maxwell counterpart the $4d$ temperature is zero at $r_+\rightarrow0$. Second, the screening effect of the charge by $\kappa$ also appears here. The nonlinearity of gravity makes the black holes appear Schwarzschild-AdS-like. As in the case of the Maxwell counterpart, this phenomenon happens only in four dimensions. 

In $D=5$, the temperature reads,
\begin{equation}
\label{T5BI}
T_{H} = -\frac{\lambda  r_+^5 \left(\sqrt{\frac{4 \kappa  q^2}{\lambda  r_+^6}+1}+1\right) \left(r_+^2 \left(\sqrt[3]{2} \lambda ^{2/3} \left(\sqrt{\frac{4 \kappa  q^2}{\lambda  r_+^6}+1}+1\right)^{2/3}-2\right)-4 \kappa \right)}{8 \pi  \kappa  \left(4 \kappa  q^2+\lambda r_+^6 \left(\sqrt{\frac{4 \kappa  q^2}{\lambda  r_+^6}+1}+1\right)\right)}
\end{equation}
\begin{figure}[htbp]
	\centering\leavevmode
	\epsfysize=6cm \epsfbox{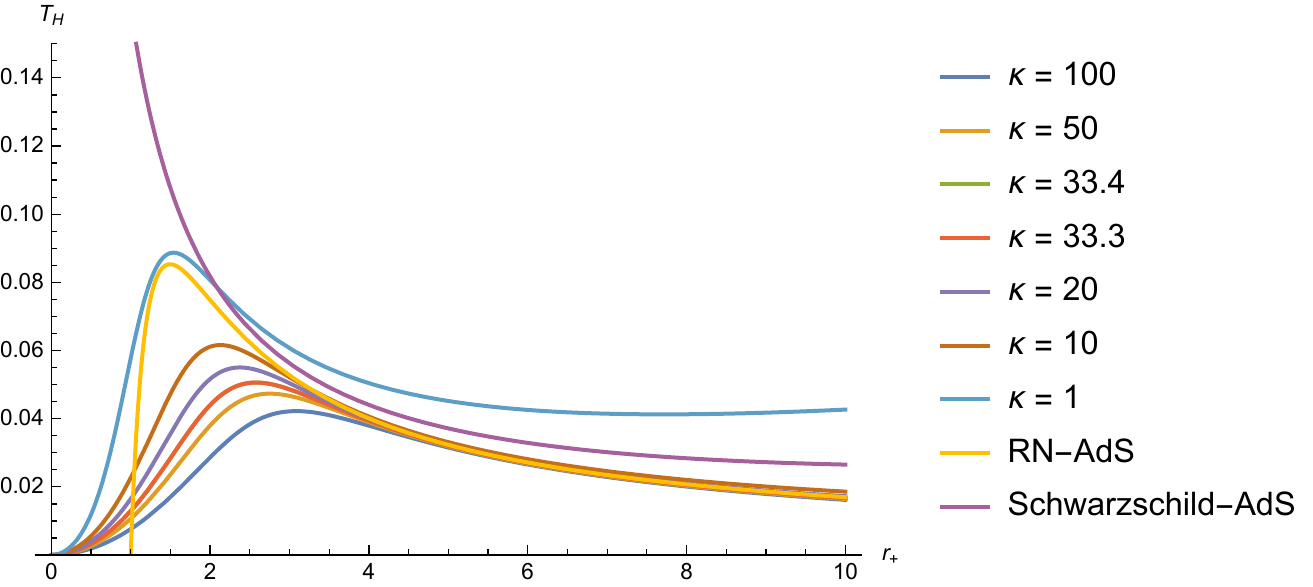}
	\caption {A plot of $5d$ EiBI-BI $T_H$  in \eqref{T5BI} with $ q=1 $ and $ \Lambda=-0.01 $.}
	\label{fig:t5bornfig}
\end{figure}  
The graph of the $5d$ temperature shown in Fig.~\ref{fig:t5bornfig}. Compare to its $4d$ temperature, there are no screening effect occurs.

As is well known, the study of BI black holes and their thermodynamical properties is  vast enough in Einstein gravity (see, for example, ~\cite{Cai:2004eh,Dey:2004yt}) but far less in EiBI gravity. In fact, to the best of our knowledge, there has been no literature on EiBI-BIthermodynamics. Here we shall show that the first law is also satisfied in EiBI-BI. To begin, we set $ e^{2\psi(\bar{r}_{+})}=0 $ at the horizon. Eq.~\eqref{67} gives
\begin{equation}
\label{1lawbi}
\left(\frac{1-V}{\kappa}\right)\bar{r}_{+}^2 = (D-3)- \bar{r}_{+} (e^{2\psi})'.
\end{equation}
The metric $V(\bar{r})$ can be expanded in powers of $\kappa$,
\begin{eqnarray}
\left(\frac{1}{\lambda} - \frac{\kappa q^2}{\bar{r}_+^{2(D-2)}}\right)^{\frac{2}{D-2}}= && \lambda ^{-\frac{2}{D-2}}-\frac{2 \kappa  \left(q^2\lambda^{1-\frac{2}{D-2}} \bar{r}_+^{-2 (D-2)}\right)}{D-2}-\frac{(D-4)\kappa ^2 q^4 \lambda^{2-\frac{2}{D-2}} \bar{r}_+^{8-4D}}{(D-2)^2} \nonumber \\ && -\frac{2 \kappa ^3 \left((D-4) (D-3) q^6\lambda^{3-\frac{2}{D-2}} \bar{r}_+^{12-6 D}\right)}{3 (D-2)^3}+O\left(\kappa ^4\right).
\end{eqnarray}
Eq.~\eqref{1lawbi} then becomes
\begin{equation}
T~ 2 \pi \bar{r}_{+}^{D-3}d\bar{r}_{+} - d\left( \frac{\bar{r}^{D-3}_+}{2} + \sum_{i=1}^{\infty} \frac{\alpha_i ~\kappa^{i-1} q^{2i}~ \lambda^{i-\frac{2}{D-2}}}{\bar{r}^{2i(D-2)+1-D}_{+}} \right) = \tilde{P}_{D} ~ d\left(\frac{\omega_{D-2}~ \bar{r}_{+}^{D-1}}{D-1} \right),
\end{equation}
where $ \tilde{P}_{D} \equiv -  \frac{\Lambda_{D}}{8 \pi \lambda} $ and $\alpha_i$ are in Eq.~\eqref{alpha}. The ADM mass is therefore 
\begin{equation}
M(\bar{r}_+)\equiv\frac{\bar{r}^{D-3}_+}{2} + \sum_{i=1}^{\infty} \frac{\alpha_i ~\kappa^{i-1} q^{2i}~ \lambda^{i-\frac{2}{D-2}}}{\bar{r}^{2i(D-2)+1-D}_{+}}.
\end{equation}
The entropy can be extracted as  
\begin{equation}
S = \int 2 \pi \bar{r}_{+}^{D-3} d\bar{r}_{+} = \frac{2 \pi \bar{r}_{+}^{D-2}}{(D-2)}. 
\end{equation} 
As expected, since we are working in $\alpha=1$, the entropy reduces to that of Tangherlini.

To check the stability, we calculate the heat capacity in eq.~\eqref{c4}. We start with the $4d$,
\begin{equation}
\label{cqbi4}
C_q\equiv \frac{\mathcal{X}(r_{+})_4}{\mathcal{Z}(r_+)_4},
\end{equation}
where
\begin{eqnarray}
\label{cqbi1}
\mathcal{X}(r_+)_4 &\equiv& \pi  \sqrt{B_4} \lambda  r_+^4 \left(4 \kappa  q^2+\lambda  r_+^4\right) \bigg[\kappa  q^2 \left(r_+^2 \left(\left(\sqrt{B_4}+3\right) \lambda -2\right)-2 \kappa \right) \nonumber \\ && +\left(\sqrt{B_4}+1\right) \lambda  r_+^4 \left((\lambda -1) r_+^2-\kappa \right)\bigg],
\end{eqnarray}
\begin{eqnarray}
\label{cqbi2}
\mathcal{Z}(r_+)_4 &\equiv& \sqrt{B_4} r_+^2 \bigg[4 \kappa ^2 q^4 \left(r_+^2 \left(\left(\sqrt{B_4}+5\right) \lambda -6\right)-2 \kappa \right) \nonumber \\ && +\kappa  \lambda  q^2 r_+^4 \left(-4 \sqrt{B_4} \kappa +r_+^2 \left(9 \sqrt{B_4} \lambda -12 \sqrt{B_4}+11 \lambda -14\right)-2 \kappa
   \right) \nonumber \\ && +\left(\sqrt{B_4}+1\right) \lambda ^2 r_+^8 \left(\kappa +(\lambda -1) r_+^2\right)\bigg],
\end{eqnarray}
and
\begin{equation}
B_4(r_{+}) \equiv \frac{4 \kappa  q^2}{\lambda  r^4_{+}}+1.
\end{equation}
\begin{figure}[htbp]
	\centering\leavevmode
	\epsfysize=6cm \epsfbox{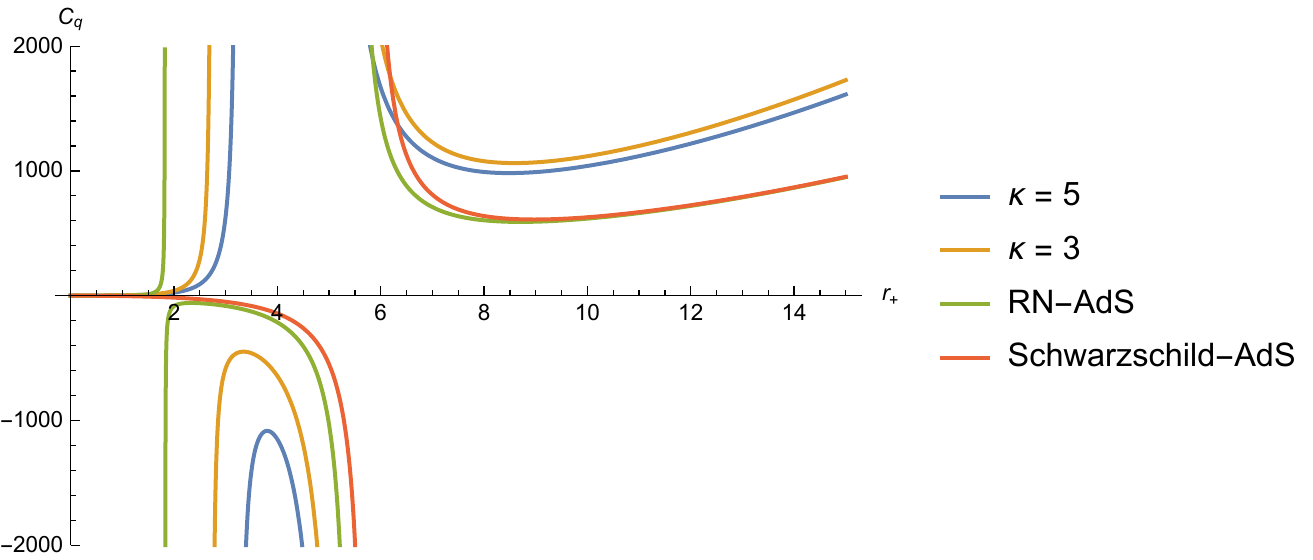}
	\caption {A plot of $4d$ heat capacity $ C_{q} $ in \eqref{cqbi4} with $ q=1 $ and $ \Lambda=-0.03 $.}
	\label{fig:gambarcbi4fig}
\end{figure}
The graph of the $4d$ heat capacity shown in Fig.~\ref{fig:gambarcbi4fig}. The negativity of the heat capacity indicates the instability of the black holes. From the Fig.~\ref{fig:gambarcbi4fig} we can infer that the nonlinearity of the gravity enables the existence of stable black holes to have smaller horizons than its RN counterpart.

In $D=5$, the heat capacity reads,
\begin{equation}
\label{cqbi5}
C_q = \frac{\mathcal{X}(r_{+})_5}{\mathcal{Z}(r_{+})_5}
\end{equation} 
where 
\begin{eqnarray}
\label{cqbi1}
\mathcal{X}(r_+)_5 &\equiv& \pi  \left(-\left(\sqrt{B_5}+1\right)^{7/3}\right) \sqrt{B_5} \lambda ^2 r_+^{10}  \left(r_+^2 \left(\sqrt[3]{2} \left(\sqrt{B_5}+1\right)^{2/3} \lambda ^{2/3}-2\right)-4 \kappa \right) \nonumber \\ && \times \left(\left(\sqrt{B_5}+1\right) \lambda  r_+^6+4 \kappa  q^2\right),
\end{eqnarray}
\begin{eqnarray}
\label{cqbi2}
\mathcal{Z}(r_+)_5 &\equiv& 96 \kappa^2 q^4 \left(4 \sqrt[3]{\sqrt{B_5}+1} \kappa +r_+^2 \left(4 \sqrt[3]{\sqrt{B_5}+1}-\sqrt[3]{2} \left(\sqrt{B_5}+4\right) \lambda ^{2/3}\right)\right) \nonumber \\ && +q^2 \bigg[\kappa  r_+^8 \left(24 \sqrt[3]{\sqrt{B_5}+1} \left(8 \sqrt{B_5}+9\right) \lambda - 12 \sqrt[3]{2} \left(13 \sqrt{B_5}  +15\right) \lambda ^{5/3}\right)  \nonumber \\ &&+48 \sqrt[3]{\sqrt{B_5}+1} \left(4 \sqrt{B_5}+3\right) \kappa ^2 \lambda  r_+^6 \bigg]
\end{eqnarray}
 and
\begin{equation}
B_5(r_{+}) \equiv \frac{4 \kappa  q^2}{\lambda  r^6_{+}}+1.
\end{equation}
\begin{figure}[htbp]
	\centering\leavevmode
	\epsfysize=6cm \epsfbox{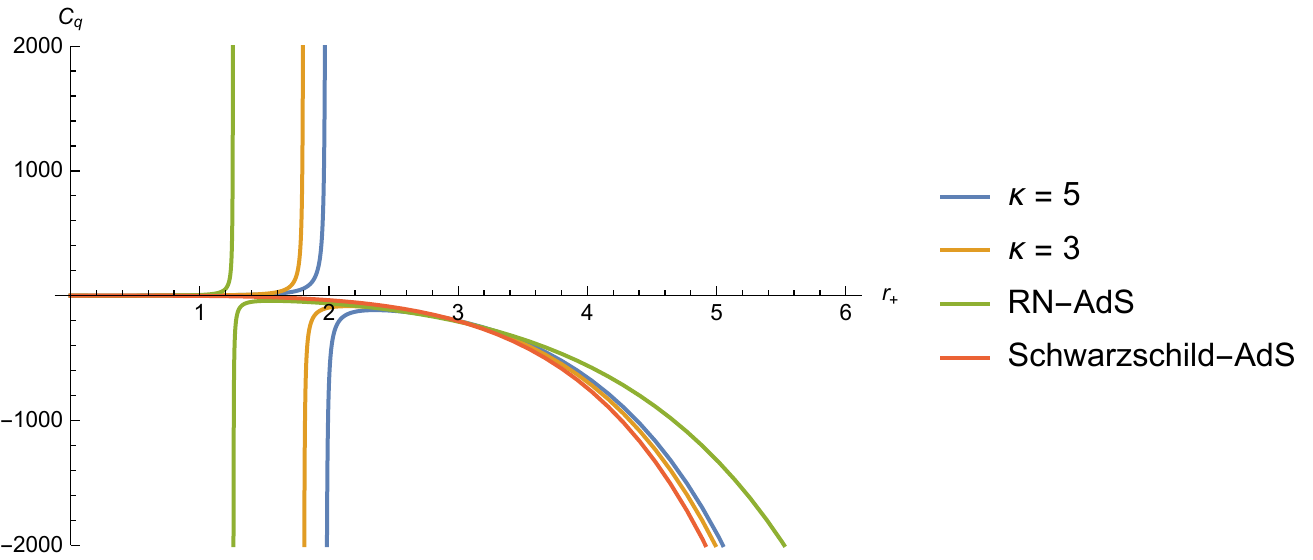}
	\caption {A plot of $5d$ heat capacity $ C_{q} $ in \eqref{cqbi5} with $ q=1 $ and $ \Lambda=-0.03 $.}
	\label{fig:gambarcbi5fig}
\end{figure}

\section{Conclusions}
\label{sec:conclu}

In this paper, we study the EiBI in $N$-dimensions coupled to $U(1)$ gauge theory, both in Maxwell as well as nonlinear electrodynamics (NLED), in particular focussing on the Born-Infeld (BI) electrodynamics. The exact solutions we obtain are the higher-dimensional generalization of the known EiBI-Maxwell and EiBI-BI black holes~\cite{Sotani:2014lua, Jana:2015cha}.

For the case of Maxwell, we found that only $4d$ electric field produced is regular. The higher-dimensional counterpart are all singular at the origin. In all dimensions we also found that the black hole singularity (inside the corresponding horizons) is surface-like. 

For the EiBI-BI case, we utilize the metric ansatz employed by Jana and Kar~\cite{Jana:2015cha}. This form enables us to reduce the field equations into algebraic ones, parametrized by the constant $\alpha\equiv4\kappa b^2/\lambda$. Unfortunately, unlike in $4d$ case, the solutions for arbitrary $\alpha$ cannot be integrated exactly. We therefore restrict our work to $\alpha=1$ and $2$. Under this constraint, all the metric and electric fields in arbitrary dimensions are solved exactly. We show that the BI electrodynamics is able to regularize the electric fields in higher dimensions. It is also possible to transform the metric solutions back into the Tangherlini gauge. As discussed in~\cite{Jana:2015cha}, the $4d$ solutions are regular at the origin. This, or course, does not mean that these black holes are of Bardeen type~\cite{Bardeen:1973gs}, since the scalar curvature are all singular. This behavior rather reflects the fact that in this theory the source charge is point-like. Our investigation reveals that the higher-dimensional metrics do not share the same property. They are all singular. 

Lastly, we study their asymptotically-AdS thermodynamics. We prove that both EibI-Maxwell and EiBI-BI black holes satisfy the first-law of thermodynamics near their horizons. The entropy extracted shows modifications from the ordinary GR-black holes. These result generalize that of~\cite{He:2016yuc} for higher dimensions. Another information extracted from the first-law is the ADM mass. It can be written as an infinite series of some transcendent function. The nonlinearity of EiBI theory imposes an effective ``screening" on the charge of the $4d$ solutions. It manifests in their Hawking temperature that abruptly switches from RN-AdS-like to Schwarzschild-AdS-like at some critical EiBI constant, $\kappa_c$. From the heat capacity, it can be inferred that the EiBI-BI black holes are be stable with smaller horizons.

We have not been able to obtain exact solutions for general $\alpha$. This work requires delicate integration technique that deserves further study. Neither do we attempt to exhaustively explore all possible NLED models. Indeed in the literature there is vast discussion on nonlinear electrodynamics toy models other than BI. In the next publication we are planning to address the EiBI black hole with different types of NLED~\cite{toappear2}.

\acknowledgments

We thank Aulia Kusuma, Ilham Prasetyo, Haryanto Siahaan, and Anto Sulaksono for enlightening discussions. This work is partially funded by Hibah PITTA UI (2267/ UN2.R3.1/HKP.05.00/2018) and Hibah Q1Q2 UI (NKB-0270/UN2.R3.1/HKP.05.00/2019).

\end{document}